\documentclass[iop,numberedappendix,appendixfloats,twocolappendix,revtex4]{emulateapj}

\usepackage{amsmath, latexsym, color, verbatim}
\usepackage{graphicx,epsfig, natbib, epstopdf}
\usepackage{pdflscape}

\def\alwaysmath#1{\ifmmode{#1}\else{$#1$}\fi}

\lefthead{Law et al. 2018}
\righthead{Spectroscopy of Faint AGN at $z\sim2$}

\slugcomment{DRAFT: \today}

\begin{document}

\newcommand{\msun}{\ensuremath{\rm M_\odot}}
\newcommand{\msunyr}{\ensuremath{\rm M_{\odot}\;{\rm yr}^{-1}}}
\newcommand{\Ha}{\ensuremath{\rm H\alpha}}
\newcommand{\Hb}{\ensuremath{\rm H\beta}}
\newcommand{\lya}{\ensuremath{\rm Ly\alpha}}
\newcommand{\Ntwo}{[\ion{N}{2}]}
\newcommand{\kms}{\textrm{km~s}\ensuremath{^{-1}\,}}
\newcommand{\ztwo}{\ensuremath{z\sim2}}
\newcommand{\zthree}{\ensuremath{z\sim3}}
\newcommand{\feh}{\textrm{[Fe/H]}}
\newcommand{\afeh}{\textrm{[$\alpha$/Fe]}}
\newcommand{\nifeh}{\textrm{[Ni/Fe]}}
\newcommand{\htwo}{\textrm{H\,{\sc ii}}}
\newcommand{\ofour}{\textrm{[O\,{\sc iiii}]}}
\newcommand{\othree}{\textrm{[O\,{\sc iii}]}}
\newcommand{\otwo}{\textrm{[O\,{\sc ii}]}}
\newcommand{\ntwo}{\textrm{[N\,{\sc ii}]}}

\newcommand{\sitwo}{\textrm{Si\,{\sc ii}}}
\newcommand{\oone}{\textrm{O\,{\sc i}}}
\newcommand{\osix}{\textrm{O\,{\sc vi}}}
\newcommand{\ctwo}{\textrm{C\,{\sc ii}}}
\newcommand{\sifour}{\textrm{Si\,{\sc iv}}}
\newcommand{\cfour}{\textrm{C\,{\sc iv}}}
\newcommand{\nfive}{\textrm{N\,{\sc v}}}
\newcommand{\mgii}{\textrm{Mg\,{\sc ii}}}
\newcommand{\fetwo}{\textrm{Fe\,{\sc ii}}}
\newcommand{\altwo}{\textrm{Al\,{\sc ii}}}
\newcommand{\hetwo}{\textrm{He\,{\sc ii}}}

\newcommand{\cgs}{\textrm{erg s$^{-1}$ cm$^{-2}$}}
\newcommand{\cgsang}{\textrm{erg s$^{-1}$ cm$^{-2}$ \AA$^{-1}$}}
\newcommand{\cgsangas}{\textrm{erg s$^{-1}$ cm$^{-2}$ \AA$^{-1}$} arcsec$^{-2}$}

\title{Imaging Spectroscopy of Ionized Gaseous Nebulae around Optically Faint AGN at Redshift $z \sim 2$}

\author{David R.~Law\altaffilmark{1}, Charles C.~Steidel\altaffilmark{2},
Yuguang Chen\altaffilmark{2}, Allison L.~Strom\altaffilmark{3}, Gwen C.~Rudie\altaffilmark{3}, Ryan F.~Trainor\altaffilmark{4}}

\altaffiltext{1}{Space Telescope Science Institute, 3700 San Martin Drive, Baltimore, MD 21218; dlaw@stsci.edu}
\altaffiltext{2}{California Institute of Technology, MS 249-17, Pasadena, CA 91125}
\altaffiltext{3}{The Observatories of the Carnegie Institution for Science, 813 Santa Barbara Street, Pasadena, CA 91107}
\altaffiltext{4}{Franklin \& Marshall College, Department of Physics and Astronomy, 415 Harrisburg Pike, Lancaster, PA 17603}

\begin{abstract}

We present Keck/OSIRIS laser guide-star assisted adaptive optics (LGSAO) integral field spectroscopy of \othree\ $\lambda 5007$ nebular emission from
twelve galaxies hosting optically faint ($\cal{R} =$ 20 - 25; $\nu L_{\nu} \sim 10^{44} -10^{46}$ erg s$^{-1}$) 
active galactic nuclei (AGN) at redshift $z \sim 2 - 3$.  In combination with deep {\it Hubble Space Telescope} Wide Field
Camera 3 ({\it HST}/WFC3) rest-frame optical imaging, Keck/MOSFIRE rest-optical spectroscopy, and Keck/KCWI rest-UV integral field spectroscopy
we demonstrate that both the continuum and emission-line structure of these sources
exhibit a wide range of morphologies from compact isolated point sources to double-AGN merging systems with extensive $\sim 50$ kpc tidal tails.
One of the twelve galaxies previously known to exhibit a proximate damped Ly$\alpha$ system coincident in redshift with the galaxy
shows evidence for both an extended \othree\ narrow-line emission region and spatially offset Ly$\alpha$ emission (with morphologically distinct blueshifted and redshifted components)
indicative of large scale gas flows photoionized by the central AGN.
We do not find widespread evidence of star formation in the host galaxies surrounding these AGN; the \othree\ velocity dispersions
tend to be high  ($\sigma = 100 - 500$ \kms), the continuum morphologies are much more compact than a mass-matched star forming comparison sample, 
and the diagnostic nebular emission line ratios are dominated by an AGN-like ionizing spectrum.
The sample is most consistent with a population of AGN that radiate at approximately their Eddington limit and photoionize extended \othree\ nebulae whose characteristic sizes scale approximately as the square root of the AGN luminosity.
\end{abstract}


\section{INTRODUCTION}
\label{intro.sec}



In the last two decades our theoretical understanding of cosmological structure formation has been shaped by the successes of the
$\Lambda$CDM model in reproducing large scale structures and the hierarchical growth of galaxy halos 
\citep[e.g.,][]{springel05,klypin11}.
While these models have been able to reproduce the observed power spectrum and merger history of the dark halos however, the 
luminous galaxies residing in their cores are a product of complicated baryonic physics describing a wide variety of physical processes 
relating to gas accretion, star formation, AGN and stellar feedback mechanisms \citep[see, e.g.,][]{hopkins18}, all of which are much less well understood. 

At no era is this deficiency more keenly felt than at redshift $z \sim 2-3$, at which both the bulk of the stellar mass in the universe is thought to have 
formed \citep[e.g.,][and references therein]{madau14}
and the space density of active QSOs appears to have peaked \citep[e.g.,][]{richards06}.
Typical star-forming galaxies at $z \sim 2-3$ are already known to exhibit ubiquitous large-scale outflows of gas \citep[e.g.,][]{steidel10,wisnioski17}
powered by the combined supernova explosions resulting from high star formation rates \citep[typically $\sim$ 30 $M_{\odot}$/yr,][]{erb06,steidel14}.
In combination with high inferred cold gas fractions of 50\% or more \citep{daddi10,papovich16} and gas-phase velocity dispersions
$\sim 60-100$ \kms\ \citep[e.g.,][]{law07,law09,fs09,wisnioski15,simons17}, conditions appear to be ripe for the onset of 
AGN formation which may be ultimately responsible for truncating the rapid star formation rates observed in the $z \sim 2-3$ galaxy sample.

In the nearby universe such AGN show a mix of broad (FWHM $\sim$ 2000 \kms) permitted and narrow (FWHM $\sim$ 500 \kms) forbidden
emission lines due to transitions arising in gas stratified by density and photoionized by the central AGN.  While the high-density broad-line regions (BLR) are extremely compact
($\ll 1$ pc) the narrow-line regions (NLR) trace low-density gas and can range in size
from a few hundred parsecs to a few kiloparsecs (extended NLR; eNLR) 
as traced by narrowband imaging \citep[e.g.,][]{mulchaey96,bennert02,sun18}, long-slit spectroscopy \citep[e.g.,][]{bennert06,rice06,greene11}, and
(more recently) integral-field spectroscopic surveys \citep[e.g.,][]{liu13,husemann14,karouzos16}.
Indeed, for particularly luminous obscured QSOs ($L_{\othree} > 10^{42}$ erg s$^{-1}$) \citet{greene11} concluded that the hard ionizing spectrum of the AGN
can effectively ionize the interstellar medium throughout the entire host galaxy.\footnote{As distinct from low-ionization nuclear emission-line region objects
(LINERS) that have similarly strong diagnostic emission line ratios but are more consistent with photoionization by hot evolved stars rather than a central source \citep[e.g.,][]{belfiore16}.}
Despite a wealth of observational data however the mechanism by which these AGN are triggered remains unclear, with evidence both
for \citep{bennert08,cotini13} and against \citep{villforth17} the significant role of major mergers.

At high redshifts multiple studies have focused on hyperluminous ($\cal{R} \sim$ 16; $\nu L_{\nu} \sim 10^{47} -10^{48}$ erg s$^{-1}$) 
broad-line QSOs \citep[e.g.,][]{canodiaz12,glikman15,bischetti17,nesvadba17},
finding evidence of substantial star formation in the host galaxies \citep[e.g.,][]{az16,carniani16} and suggesting that while AGN feedback may quench star formation in
some regions of the galaxy this quenching may be incomplete \citep[or indeed, trigger enhanced star formation near the edges of the outflows;][]{cresci15}.
Although large IFU surveys \citep[e.g.,][]{fs09,fs18} serendipitously contain some fainter narrow-line 
AGN \citep[especially at stellar masses $M_{\ast} > 10^{11} M_{\odot}$; see, e.g.,][]{fs14,genzel14}, it is only recently that 
targeted spatially resolved \citep[e.g.,][]{harrison16,kakkad16} and long-slit spectroscopy  \citep[e.g.,][]{azadi17,leung17} of this population
has become more common, and samples are still limited to lower redshifts and/or lower spatial resolution.
In this contribution we extend the analysis of the spatially resolved properties of AGN eNLR to redshifts $z \sim 2-3$
by combining  {\it Hubble Space Telescope} ({\it HST}) WFC3 rest-optical imaging with observations
from the Keck OSIRIS \citep{larkin06}, MOSFIRE \citep[][]{mclean12}, LRIS \citep[][]{oke95,steidel04}, KCWI \citep{morrissey12}, NIRSPEC \citep{mclean98}, and ESI \citep{sheinis02} spectrographs.
We examine the resolved spatial structure of nebular \othree\ $\lambda 5007$ ionized 
gas emission surrounding a sample of optically faint ($\cal{R} =$ 20 - 25; $\nu L_{\nu} \sim 10^{44} -10^{46}$ erg s$^{-1}$) 
AGN identified on the basis of high-ionization rest-UV emission line features
whose broadband colors and spatial clustering properties are consistent with those of the star forming galaxy population \citep{steidel02,adel05}.
Through these observations we assess 1) How the spatial morphology (continuum and spectral line emission) of the AGN host galaxies compares to
the star forming parent population, and whether they typically show hallmarks of recent major mergers,  2) Whether we can detect and characterize star formation within the host galaxies or identify
spatially resolved outflows driven by the central AGN,  and 3) Whether the extent of the AGN-photoionized eNLR conforms to expectations based
on observations of low-redshift sources.

In \S \ref{sample.sec} we discuss our sample selection and the wide variety of ancillary imaging and spectroscopic data available for our targets.
\S \ref{osirisobs.sec} describes our OSIRIS observing program, data reduction strategy, and typical data quality.
We discuss our observations of individual galaxies in \S \ref{results.sec}, and the interpretation of these observations for the 
$z \sim 2$ AGN population in  \S \ref{discussion.sec}.
We summarize our conclusions in \S \ref{conclusions.sec}.
Throughout our analysis we assume a  $\Lambda$CDM cosmology 
in which $H_0 = 70$ km s$^{-1}$ Mpc$^{-1}$, $\Omega_{\rm m} = 0.27$, 
and $\Omega_{\rm \Lambda} = 0.73$.

\section{Sample Selection and Ancillary Data}
\label{sample.sec}

\subsection{Sample Selection}

\begin{figure}
\epsscale{1.2}
\plotone{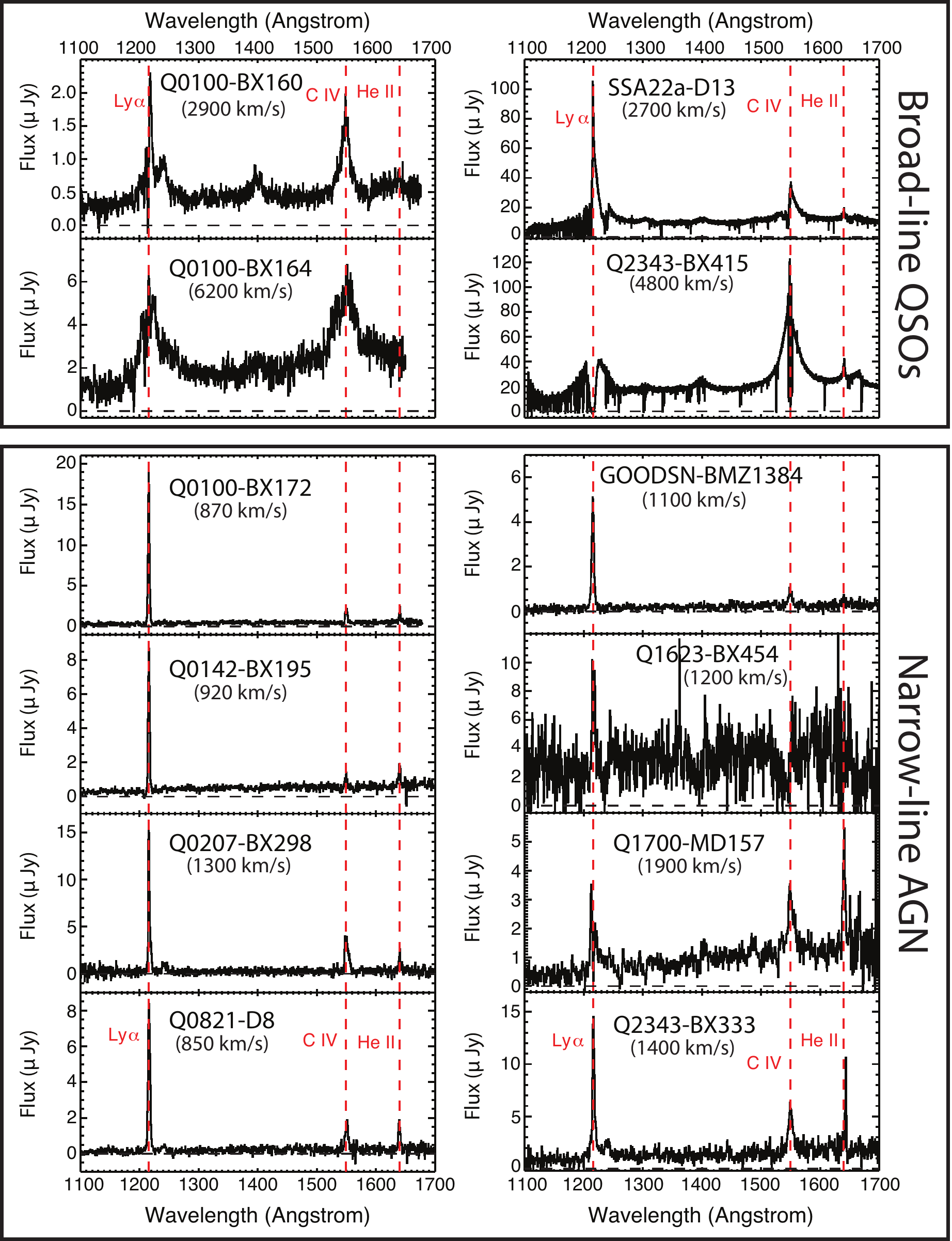}
\caption{Keck/LRIS and Keck/ESI (SSA22a-D13 and Q2343-BX415) rest-UV spectra of the $z \sim 2-3$ AGN sample.  
All twelve systems show Ly$\alpha$,  \cfour\ 1550, and \hetwo\ in emission while some (most clearly Q0100-BX160) also show \nfive\ 1240 emission on the wing of the \lya\ profile.
Although there are a wide range of profiles, four are formally classified as broad-line systems while eight are classified as narrow-lined systems.  Inset text in each panel gives the observed
FWHM of the \cfour\ feature; since the profile for SSA22a-D13 is asymmetric the FWHM was obtained by doubling the half-width at half maximum (HWHM) from the red side of the line profile.
}
\label{uvspec.fig}
\end{figure}

We identified a sample of 12 galaxies (Table \ref{targets.table}) containing faint AGN in the redshift range $z = 2.0 - 3.4$ from the
Keck Baryonic Structure Survey  \citep[KBSS, codified as such by][]{rudie12,ts12} on the basis of high-ionization rest-UV emission features
present in Keck/LRIS and/or Keck/ESI spectroscopy.  In particular, we select objects with
Ly$\alpha$, \cfour\ 1550, and \hetwo\  detections (often accompanied by \nfive\ 1240 as well) since typical $z\sim2$ star forming galaxies do not have a 
sufficiently hard ionizing spectrum to produce emission in these lines.
As illustrated in Figure \ref{uvspec.fig}, these 12 galaxies exhibit 
\cfour\ 1550 emission lines ranging in width from 850 to about 6200 \kms\ FWHM that we divide into two general categories; 
broad-lined `Type I' QSOs with Lorentzian line profiles \citep[consistent with turbulent motions in the accretion disk; see, e.g.,][]{kollat13},
and narrow-lined `Type II' AGN whose line profiles are narrower and more gaussian in form.   

The dividing line between broad-line and narrow-line categories in the literature is not well-determined, and criteria range from 1000 to 2000 \kms\ FWHM, applied to either UV or optical emission lines depending on the wavelength range of a given study.  \citet{domin18} for instance used a threshold of 1000 \kms\ in a recent study of low-redshift AGN observed as a part of the SDSS-IV/MaNGA IFU galaxy survey \citep{bundy15,law16}, while \citet{hao05} adopt 1200 \kms\ for AGN in SDSS single-fiber spectroscopy, and \citet{steidel02} and \citet{greene14} use 2000 \kms\ for AGN and quasars
at redshifts $z > 2$.  For consistency with prior KBSS publications we adopt the \citet{steidel02} division in which 
eight of our objects have \cfour\ FWHM less than 2000 \kms\ and are thus classified as narrow-line AGN, while four are broad-lined QSOs, one
of which (Q2343-BX415) is known to exhibit a proximate damped Ly$\alpha$
absorption system (PDLA) at approximately the quasar redshift \citep[][]{rix07}.

As discussed by \citet{steidel02}, the composite spectra of the narrow-line AGN indicate that they may be similar to
local Seyfert 2 type galaxies.
Such AGN represent about 3\% of the KBSS spectroscopic sample whose redshift distributions, clustering properties,
characteristic continuum magnitudes\footnote{Generally the bright
\lya\ emission feature has negligible impact on the optical color selection since it falls between the $U_n$ and $G$ filters for the majority of our sample; see discussion by
\citet{reddy08} and \citet{hainline11}.}, and mass-matched star formation rates (SFR) derived from stellar population models
overlap those of the ordinary star forming galaxies \citep[see, e.g.,][]{steidel02, hainline12},
suggesting that they may represent an evolutionary 
phase in the lifetimes of ordinary star forming galaxies in the young universe ($z \sim 2 - 3$).
However, the AGN fraction rises significantly as a function of the stellar mass of the host galaxy \citep[][]{hainline12} resulting in a sample that is 
relatively massive ($M_{\ast} > 10^{10} M_{\odot}$) compared to the overall KBSS parent sample,
due either to an intrinsic preference for AGN formation in higher mass galaxies \citep[e.g.,][]{kauffmann03} or to incompleteness in the UV emission line selection method (since \cfour\ may be too weak to detect in the KBSS rest-UV spectra for AGN in lower mass galaxies).
At the same time, we note that the KBSS AGN selection technique tends to be more sensitive to low-luminosity AGN than X-ray selection methods and nearly as sensitive as {\it Spitzer}
mid-IR photometric techniques (which can also include dustier AGN missed by the UV selection method) as detailed by \citet{reddy06} \citep[see also][]{azadi17}.

In contrast, the spectra of the broad-line QSOs in our sample are more akin to hyperluminous QSOs \citep[e.g.,][]{ts12}, albeit
with much fainter optical magnitudes. These may not simply be the less-obscured counterparts of the narrow-line AGN since their detailed UV spectral features have a larger impact on
their observed colors (scattering them into and out of the KBSS selection boxes), and they appear to be more strongly clustered than the star-forming galaxy sample \citep[][]{steidel02}.
We therefore treat the broad-line and narrow-line samples separately in many of our analyses in cases where the AGN is expected to contribute significantly to the observed optical continuum.

In addition to the sample selection biases thus imposed by the initial identification of these galaxies in the KBSS
(i.e., rest-UV emission features, and optical colors/magnitudes matching the $z \sim 2-3$ star forming galaxy population), 
additional practical selection criteria that limit the list of available targets were adopted 
in order to optimize the likelihood of detection in our 
Keck/OSIRIS observations.  First, we required that the redshift was such that \othree\ $\lambda 5007$ emission
would avoid extremely strong atmospheric OH emission features and telluric absorption bands.  Second, we required that the targets
had a viable tip-tilt reference star for the laser guided adaptive optics (LGSAO) system within 70''; although this is not challenging in most of the KBSS
survey fields (which are centered on lines of sight to bright background quasars), it poses particular limitations in
the GOODS-N field.  Finally, since many of our observations were obtained as filler and/or backup targets during different observing
programs our sample is additionally restricted in R.A. by the available observing windows in which we were able to observe
during fair-good observing conditions (i.e., clear with optical seeing $\leq 1$ arcsec).


\begin{deluxetable*}{lccccccccccc}
\tablecolumns{12}
\tablewidth{0pc}
\tabletypesize{\scriptsize}
\tablecaption{OSIRIS Observations and Target AGN}
\tablehead{
\colhead{Name} &  \colhead{R.A.} & \colhead{Decl.}  & \colhead{$z_{\rm neb}$\tablenotemark{a}} & \colhead{Type\tablenotemark{b}} & \colhead{${\cal R}_{\rm AB}$} & \colhead{Date} & \colhead{Time\tablenotemark{c}} & \colhead{Filter} & \colhead{Scale} & \colhead{Line} & \colhead{PA\tablenotemark{d}} \\
 & \colhead{(J2000)} & \colhead{(J2000)} & & & & & & & (mas) }
\startdata
\multicolumn{12}{c}{Broad-line QSOs}\\
\hline
Q0100-BX160 & 01:03:07.542 & +13:17:02.57            & 2.3032         & QSO & 24.43 & Sep 2008 & 0.25 & Kn3 & 50 & \Ha & 0\\
Q0100-BX164  & 01:03:07.668   & +13:16:30.93         & 2.2924 & QSO & 23.61 & Sep 2008 & 1.0 & Kn3 & 50 & \Ha  & 90\\
SSA22a-D13 &  22:17:22.259 & +00:16:40.63             & 3.3538        & QSO & 20.84 & Sep 2008 & 2.75 & Kn3 & 50 & \othree  & 236\\
 & & & & & & Jun 2009 & 2.0 & Kn2 & 50 & \Hb  & 236\\ 
Q2343-BX415 &  23:46:25.427  &  +12:47:44.32         & 2.5741        & QSO & 20.22 & Sep 2007 & 1.5 & Hn5 & 50 & \othree  & 120\\ 
& & & & & & Jun 2008  & 1.0 & Kn5 & 50 & \Ha  & 120\\ 
& & & & & & Oct 2016  & 2.75 & Hn5 & 50 & \othree  & 120\\ 
\hline
\multicolumn{12}{c}{Narrow-line AGN}\\
\hline
Q0100-BX172  &01:03:08.452  &  +13:16:41.74          & 2.3119 & AGN & 23.50 & Sep 2008 & 3.5 & Kn3 & 50 & \Ha  & 110\\
& & & & & & Sep 2008 & 1.0 & Hn3 & 50 & \othree  & 110\\ 
& & & & & & Jan 2017 & 1.5 & Hn3 & 50 & \othree  & 290\\ 
Q0142-BX195  & 01:45:17.709  &  -09:44:54.43          & 2.3807\tablenotemark{e} & AGN & 23.56 & Oct 2016 & 3.0 & Hn4 & 100 & \othree  & 0\\ 
& & & & & & Sep 2008 & 1.25 & Kn4 & 50 & \Ha  & 180\\ 
& & & & & &  Aug 2011 & 1.0 & Kc4 & 100 & \Ha  & 0 \\ 
Q0207-BX298 & 02:09:54.046 & -00:04:29.61             &  2.1449 & AGN & 25.07 & Jan 2017 & 2.0 & Hn2 & 50 & \othree  & 0\\
 & & & & & & Jan 2017 & 0.5 & Kn2 & 100 & \Ha  & 0\\ 
Q0821-D8 & 08:20:59.251 & +31:08:57.64                  & 2.5674 & AGN & 24.76 & Jan 2017 & 3.5 & Hn5 & 100 & \othree  & 0\\ 
GOODSN-BMZ1384 & 12:37:23.106 & +62:15:37.76  & 2.2428       & AGN & 23.98 & Jan 2017 & 3.0 & Hn3 & 100 & \othree  & 0\\
 & & & & & & Jan 2017 & 0.5 & Kn3 & 100 & \Ha  & 0\\ 
Q1623-BX454 &  16:25:51.417   & +26:43:46.39         & 2.4181\tablenotemark{f} & AGN & 23.89 & Sep 2008 & 0.5 & Kn4 & 50 & \Ha  & 225\\
Q1700-MD157 & 17:00:52.178 & +64:15:29.33           & 2.2928        & AGN & 24.35 & Jun 2009 & 0.25 & Kn3 &  50 &\Ha  & 290\\
Q2343-BX333 & 23:46:21.506 & +12:47:03.14            & 2.3948          & AGN & 24.12 & Sep 2008 & 0.5 & Kn4 & 50 & \Ha  & 236 
\enddata
\label{targets.table}
\tablenotetext{a}{Nebular emission line redshift from narrow line Keck/OSIRIS observations (if possible), Keck/NIRSPEC, or Keck/MOSFIRE.}
\tablenotetext{b}{`QSO' are those systems with broad (FWHM $> 2000$ km/s) rest-UV emission lines, while `AGN' are those systems with narrow emission lines.}
\tablenotetext{c}{Total OSIRIS observing time in hours.}
\tablenotetext{d}{Position angle of the OSIRIS IFU (degrees E from N).}
\tablenotetext{e}{Interacting double-AGN system; $z = 2.3807$ is the redshift of the entire core+tail system, individual AGN have redshifts $z = 2.3805$ and $z = 2.3771$.}
\tablenotetext{f}{Close pair system; companion has $z = 2.4200$.}
\end{deluxetable*}

\subsection{Ancillary Data}
\label{ancillary.sec}

\subsubsection{Imaging and Stellar Population Models}
\label{sed.sec}

\begin{figure*}[p!]
\plotone{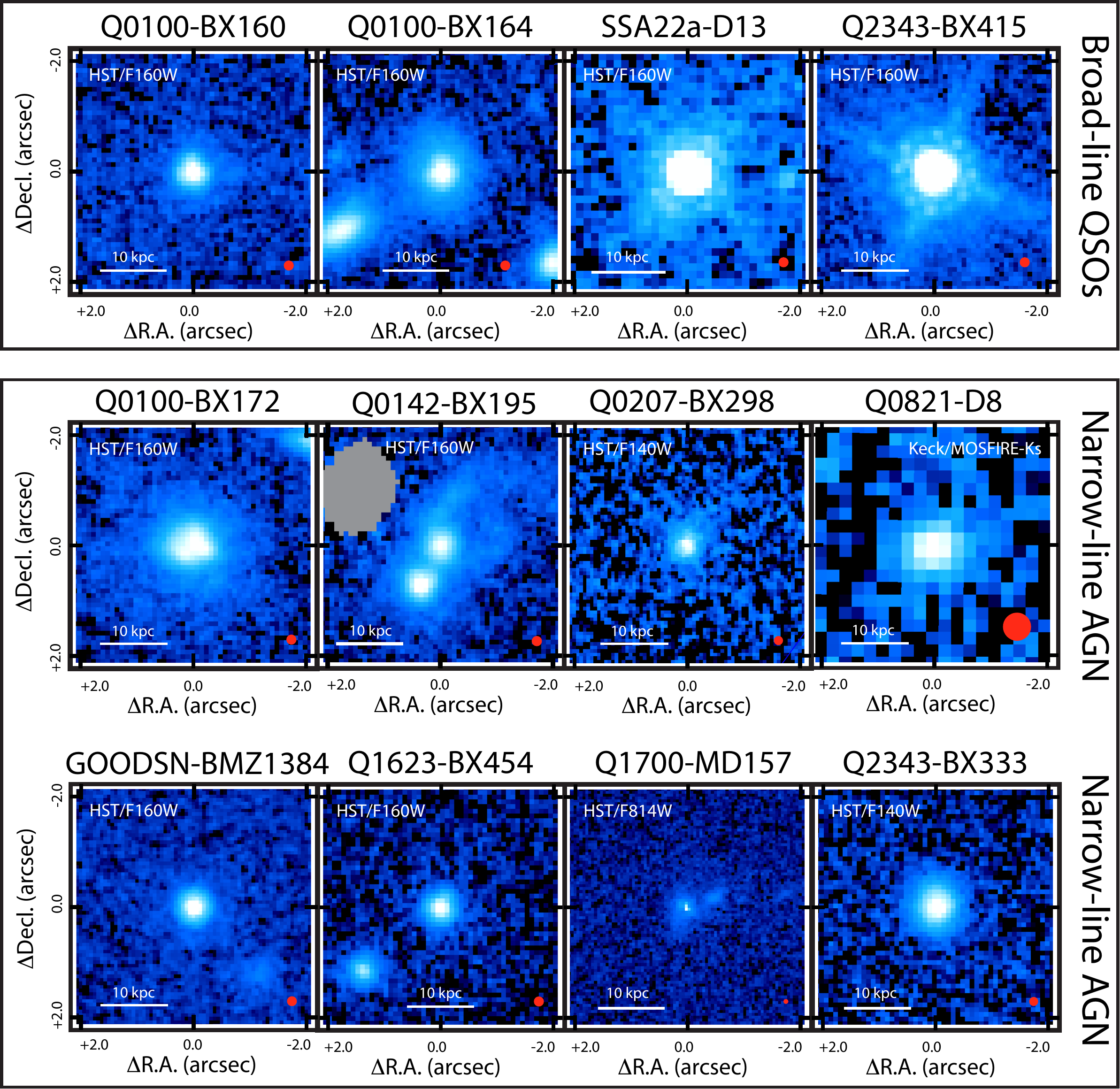}
\caption{Rest-optical or rest-UV (for Q1700-MD157) continuum morphology of the eight narrow-line and four broad-line AGN; all images except Q0821-D8 share a common
logarithmic stretch.  In each panel the red circle indicates the FWHM of the observational PSF.}
\label{hst_all.fig}
\end{figure*}

In addition to the $U_n G {\cal R}$ imaging \citep{adelberger04, steidel04} that defines the optical colors used for the initial selection of our galaxy sample all of our fields also have 
ground-based near-IR $J+K_{\rm s}$ band imaging and {\it Spitzer} IRAC photometry \citep[see, e.g.,][]{reddy12}
that we augment with {\it Hubble Space Telescope} ({\it HST}) imaging tracing the rest-frame optical galaxy morphology (Figure \ref{hst_all.fig}).
The majority of galaxies use {\it HST} WFC3/F160W imaging obtained as part of GO-11694 (PI: Law)
and processed as decribed by \citet{law12a}; additional data for Q0207-BX298 are drawn from public WFC3/F140W grism preimaging (GO-12471, PI: Erb),
for SSA22a-D13 from public WFC3/F160W data (GO-11636, PI: Siana), for Q2343-BX333 from WFC3/F140W imaging in GO-14620 (PI: Trainor),
for Q1700-MD157 from public ACS/F814W imaging (GO-10581, PI: Shapley), and for GOODSN-BMZ1384 from the CANDELS deep WFC3/F160W imaging program
 \citep{grogin11,koekemoer11}\footnote{Available online at https://candels.ucolick.org/data\_access/GOODS-N.html}.
 Photometric data points for GOODSN-BMZ1384 were taken
directly from the 3D-HST photometry catalog \citep{brammer12,skelton14}.
 No {\it HST} imaging is available for Q0821-D8 and we use Keck/MOSFIRE $K_s$-band imaging for morphological information instead.
 In each case for which \othree\ emission falls within a filter we correct the observed magnitude for the corresponding flux.

\begin{figure*}
\epsscale{1.0}
\plotone{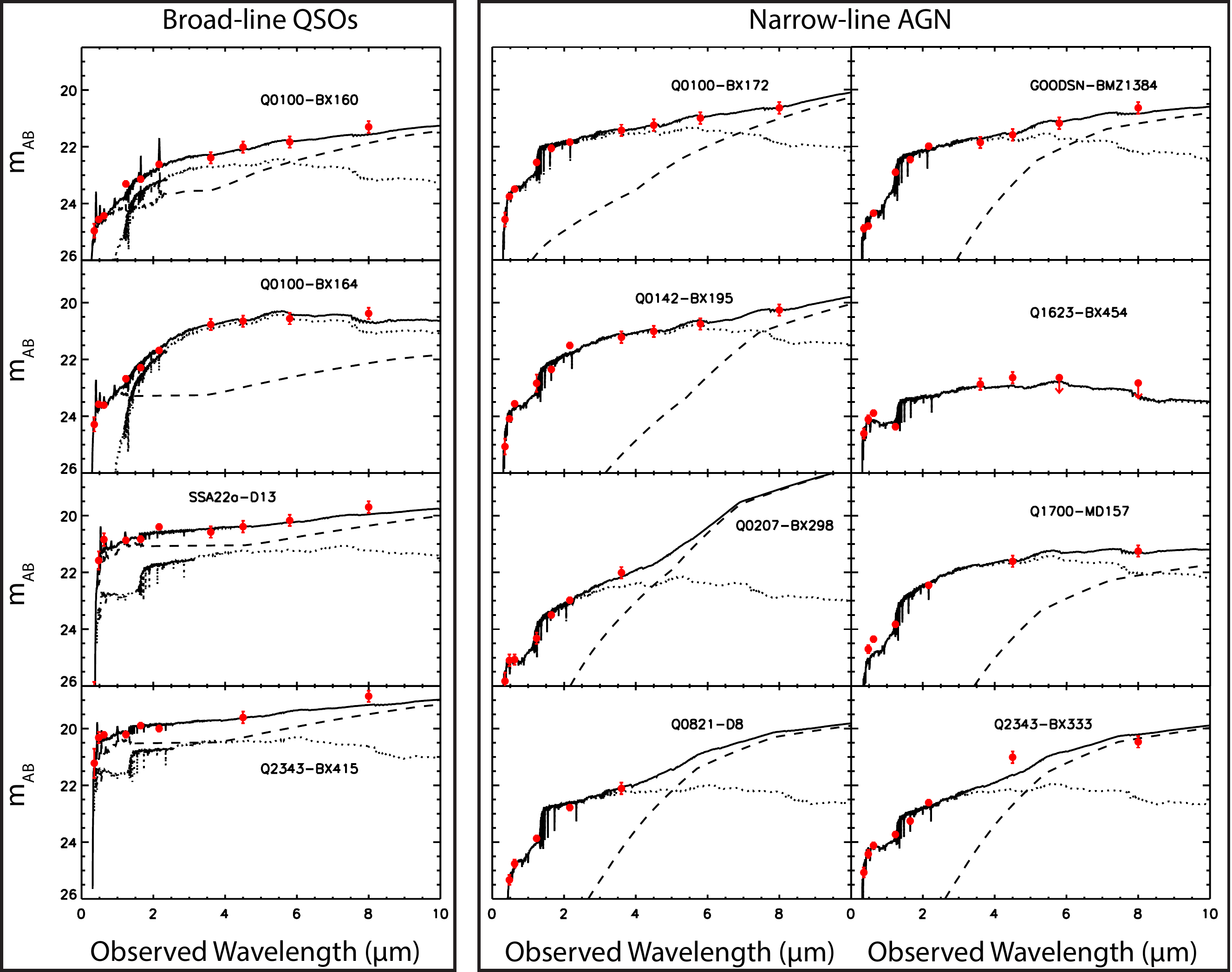}
\caption{Best-fit stellar population plus AGN template models (solid black line) overplotted against ground-based
$U_n G {\cal R} J  H K$ and {\it Spitzer}/IRAC photometry for the twelve target AGN.  
Upper limits on the observed flux densities (Q1623-BX454) are shown as downward-pointing red arrows.
By construction, the stellar population models (dotted lines) dominate the SED at short wavelengths for the narrow-line objects while the AGN models (dashed lines) provide
a good match to the rising NIR photometry.}
\label{sed.fig}
\end{figure*}

As illustrated by Figure \ref{sed.fig}, the SED of our galaxies continues to rise into the mid-infrared due to the thermal contribution from the AGN,
and simplistic stellar models are poor fits to the long-wavelength observational data.
We therefore use the FAST synthetic template fitting code \citep{kriek09,kriek18,aird18} to model the observational data as a combination of stellar and 
AGN templates.
The stellar population templates are drawn from the Flexible Stellar Population Synthesis \citep[FSPS][]{conroy10} library and adopt a
\citet{chabrier03} IMF, a \citet{calzetti00} extinction law, and an exponentially declining star formation history with fixed solar metallicity.  The AGN templates adopted vary between
our broad-line and narrow-line sources according to the expected contribution of the AGN to the optical continuum flux.

For our four broad-line objects (in which the AGN contributes significantly
to the total optical flux) we use the unobscured Torus, QSO1, QSO2, TQSO1, and BQSO1 models from the
SWIRE template library \citep{polletta07} as these span a wide range 
of IR/optical flux ratios \citep[see discussion in Appendix A of][]{aird18}.  The results, however, are degenerate in the relative contributions of AGN and stellar components to the total
optical continuum luminosity; almost equally-good fits can be obtained for SSA22a-D13 and Q2343-BX415 with AGN-dominant or stellar-dominant models depending on the assumptions
that are made about the age of the stellar population.  We therefore do not attempt to estimate the stellar mass, SFR, etc of the host galaxy for the broad-line objects as such estimates
would be highly uncertain.

In contrast, for the eight narrow-line objects in our sample we use the optically obscured AGN templates provided by \citet{silva04}.
These templates have absorption column densities $N_{\rm H} = 10^{22-23}, 10^{23-24}$, and $10^{24-25}$ cm$^{-2}$; although the majority of our galaxies are best fit with the
Compton thick AGN template with column
density $N_{\rm H} = 10^{24-25}$ cm$^{-2}$, lower density templates also provide fits that are almost indistinguishable in terms of their $\chi^2$.
The combination of these templates with the star forming stellar population component satisfactorily reproduces the observed photometry for all eight of our narrow-line objects, and we therefore
derive stellar masses, star formation rates, and $E(B-V)$ color excesses that
range from $\sim 10^{10} - 10^{11} M_{\odot}$, $4 - 34 M_{\odot}$ yr$^{-1}$, and $0-0.16$ respectively (see Table \ref{sed.table}).  These values are relatively
insensitive to the details of our approach to the modeling; our
masses are consistent with those calculated using just a stellar component with constant star formation to within 0.2 dex on average,
and for the subset of six narrow-line AGN in common with the prior study of \citet{hainline12} the masses agree to within an average of 0.1 dex.

 As discussed in depth by \citet{theios18}, for star-forming galaxies the nebular extinction as measured from the Balmer decrement (BD) tends to be higher than that derived from the
 color excess and SED modeling, indicating
 that the attenuation in the line-emitting regions is systematically higher than in the rest of the galaxy
 (albeit with large scatter).  In the present case, however, this relation might not be expected to hold since the majority of the nebular emission comes not from individual star-forming \htwo\ regions but from
 clouds of gas in the NLR photoionized by the radiation from the central AGN (see \S \ref{ratios.sec}).  Since not all of the objects in our sample have both the \Ha\ and \Hb\ observations necessary
 to measure the BD directly, in absence of a well-established relation we therefore assume that the SED-derived extinction estimates are representative of the extinction throughout the NLR 
(since both are distributed on similar physical scales) and use it to correct the nebular emission line luminosities of narrow-line AGN for dust extinction accordingly.


 
\begin{deluxetable*}{lccccc}
\tablecolumns{6}
\tablewidth{0pc}
\tabletypesize{\scriptsize}
\tablecaption{Stellar Parameters for Narrow-line AGN from Multi-Component Population Models}
\tablehead{
\colhead{Name} &  \colhead{$E(B-V)$} & SFR ($M_{\odot}$ yr$^{-1}$) & \colhead{log $(M_{\ast}/M_{\odot})$} & \colhead{log($N_{\rm H}$)\tablenotemark{a}} & \colhead{$f_{\rm AGN}\tablenotemark{b}$}
}
\startdata
Q0100-BX172 & 0.13 & 25 &  10.60 & 22.5 & 0.10 \\
Q0142-BX195 & 0.16 &  34 &10.99 & 23.5 & 0.02 \\
Q0207-BX298 & 0.06 &  4 &10.41 & 23.5 & 0.19 \\
Q0821-D8 &  0.16 &  7 &10.45 & 24.5 & 0.14 \\
GOODSN-BMZ1384 & 0.00 &  4 &10.60 & 24.5 & 0.04\\
Q1623-BX454 & 0.00 &  8 &10.09 & 24.5 & 0.00 \\
Q1700-MD157 & 0.16 &  10 &10.68 & 24.5 & 0.02 \\
Q2343-BX333 & 0.10 &  15 &10.48 & 24.5 & 0.12
\enddata
\tablenotetext{a}{Absorption column density (cm$^{-2}$) for the best-fit obscured AGN template.}
\tablenotetext{b}{Fractional contribution of the AGN to the integrated light at rest-frame 1\micron.}
\label{sed.table}
\end{deluxetable*}

\subsubsection{Spectroscopy}

\begin{figure}
\epsscale{1.2}
\plotone{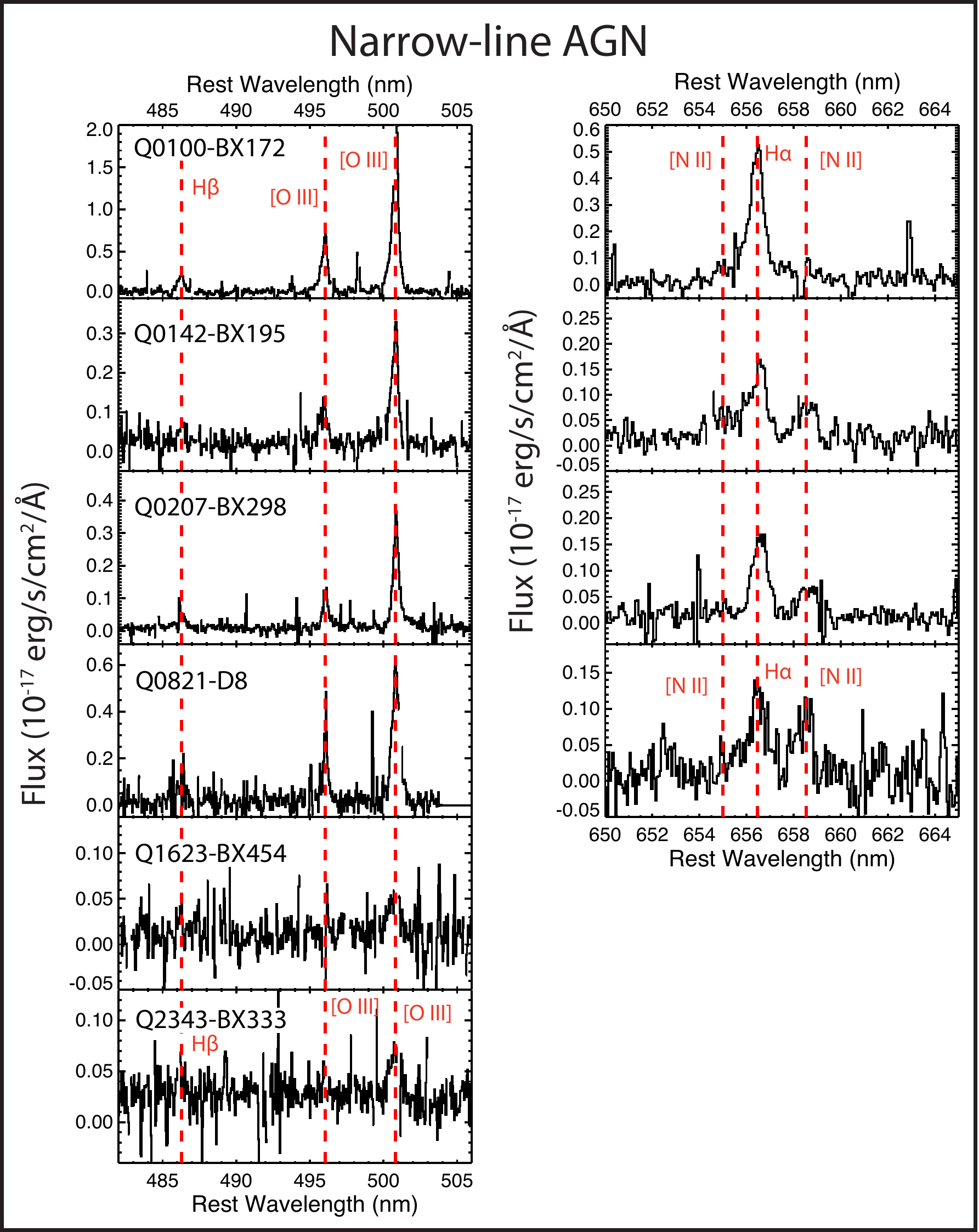}
\caption{Keck/MOSFIRE $H$ and $K$-band
spectra of the narrow-line AGN shifted to the rest frame (no slit-loss corrections have been applied).
Vertical dashed lines denote the expected locations of nebular \othree, \ntwo, \Ha, and \Hb\ emission
at the systemic redshift.  Data have been masked at the wavelengths of strong atmospheric OH emission in order to suppress
sky subtraction residuals.
}
\label{mosfire1.fig}
\end{figure}

\begin{figure}
\epsscale{1.2}
\plotone{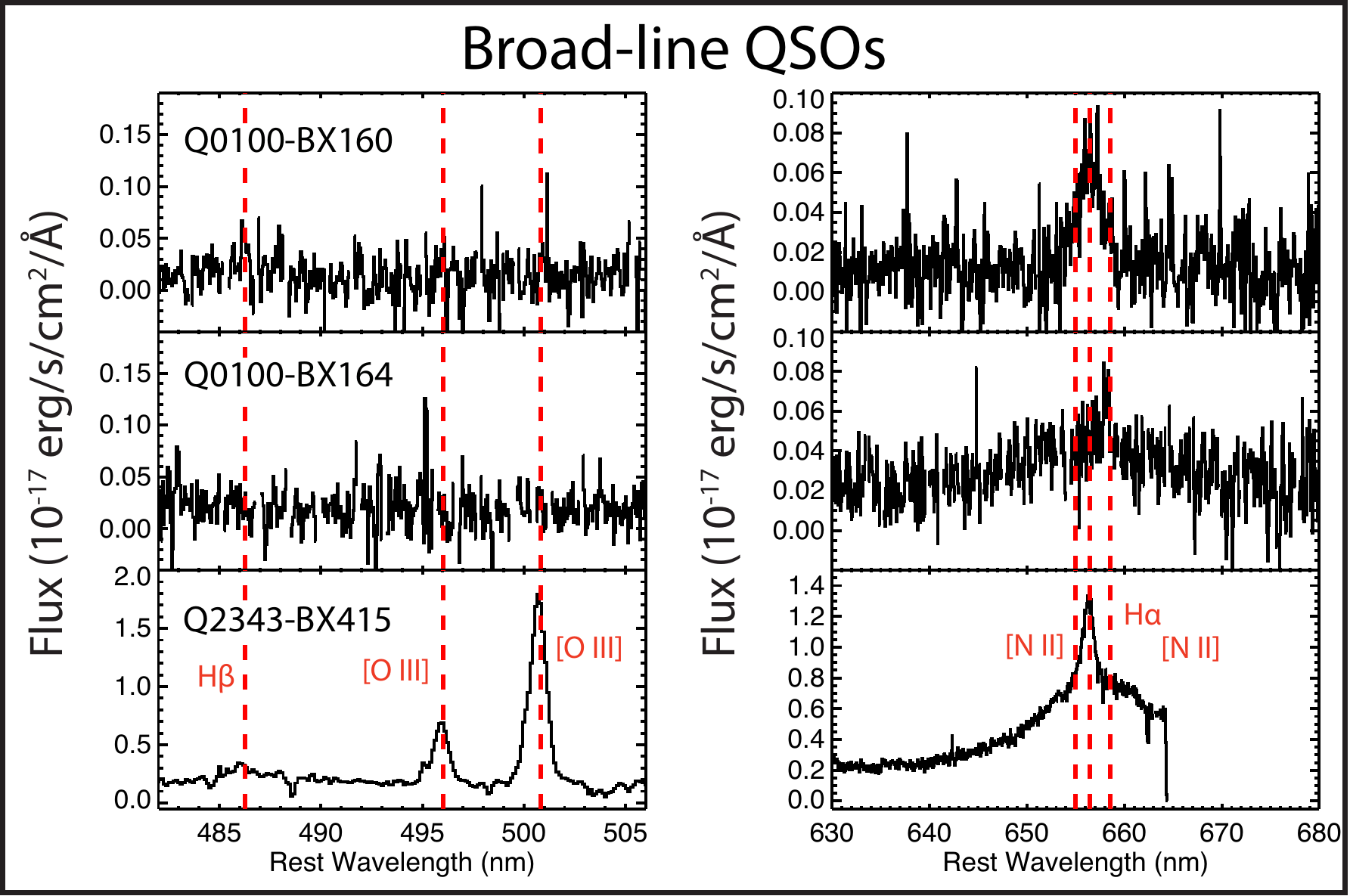}
\caption{As Figure \ref{mosfire1.fig} but for the broad-line QSOs.  Q0100-BX160 and Q0100-BX164 spectra were observed with Keck/MOSFIRE
while Q2343-BX415 was observed with Keck/NIRSPEC.
}
\label{mosfire2.fig}
\end{figure}

In addition to the Keck/LRIS optical spectroscopy \citep{shapley03,steidel10} (and in the case of SSA22a-D13 and Q2343-BX415, ESI high resolution optical spectroscopy)
that characterizes the KBSS, the majority of AGN in our study also have $R \sim 3600$
multi-band $JHK$ rest-optical Keck/MOSFIRE slit spectroscopy \citep{steidel14,strom17} covering the \otwo\ $\lambda3727$ to \Ha\ emission features.
The MOSFIRE data were obtained as a part of the larger KBSS observing program, and details of the observations and data reduction have been extensively
described by \citet{steidel14}.  As illustrated by Figures \ref{mosfire1.fig} and \ref{mosfire2.fig}, our galaxies are typically well-detected in both \othree\ and \Ha\, although a subset of the broad-line objects
shows no measurable \othree\ emission even at the limiting depth of MOSFIRE (see discussion in \S \ref{accretion.sec}).

More recently, we obtained Keck/KCWI rest-UV integral field spectroscopy of the Q0142 and Q2343 fields covering Q0142-BX195 and Q2343-BX415 respectively
on the nights of 22-23 September 2017.

The KCWI medium-scale slicer samples a contiguous field of 16\farcs{5} $\times$ 20\farcs{4}, where the longer dimension is
along slices, with $24 \, \times$ 0\farcs{69} samples perpendicular to the slices. The E2V 4k$\times$4k detector was binned 
2$\times$2 on readout, resulting in spatial sampling on the detector of 0.29'' along slices and 2.5 pixels per spectral resolution element.
The KCWI-B ``BL'' grating was used with a camera angle placing 4500 \AA\ at the center of the detector; the common wavelength
range sampled by all 24 slices is 
$3530-5530$ \AA\ ($\sim 100 - 160$ nm restframe at redshift $z \sim 2$), with a spectral resolution of 2.5~\AA\ (FWHM), 
for a resolving power of $R \sim 1400-2200$ depending on wavelength. For both Q2343-BX415 and Q0142-BX195, individual 
exposures of 1200~s each (9x1200s for BX415 and 6x1200s  for BX195) were obtained
with small telescope offsets between each to better-sample the spatial PSF in the direction perpendicular to slices. The KCWI Data Extraction and Reduction
Pipeline ([KDERP] https://github.com/Keck-DataReductionPipelines/KcwiDRP)  was used to reduce raw CCD frames to wavelength calibrated, spatially rectified, differential
atmospheric dispersion-corrected, background subtracted,
flux calibrated data cubes with inital sampling of
0.29'' by 0.69'' (spatial) and 1\AA\ pix$^{-1}$ (spectral). The individual reduced data cubes were then averaged with inverse variance
weighting after shifting into registration and re-sampling  
spatially onto an astrometrically-correct rectilinear grid sampled with 0.3'' in each spatial dimension, rotated to place North up and East left.

For convenience, we summarize our spectroscopic observations for each of our twelve galaxies in Table \ref{specsummary.table}.

\begin{deluxetable*}{lccccccccc}
\tablecolumns{10}
\tablewidth{0pc}
\tabletypesize{\scriptsize}
\tablecaption{Spectroscopic Observations Summary}
\tablehead{
\colhead{Name} &  \colhead{OSIRIS-\othree} & \colhead{OSIRIS-\Ha} & \colhead{LRIS} & \colhead{ESI} & \colhead{NIRSPEC-$H/K$} & \colhead{MOSFIRE-$J$} & \colhead{MOSFIRE-$H$} & \colhead{MOSFIRE-$K$} & \colhead{KCWI}
}
\startdata
\multicolumn{10}{c}{Broad-line QSOs}\\
\hline
Q0100-BX160 & ... & X\tablenotemark{a} & X & ... & ... & X & X & X & ...  \\
Q0100-BX164 & ... & X\tablenotemark{a} & X & ... & ... & ... & X & X & ...  \\
SSA22a-D13  & X & X\tablenotemark{b} & ... & X & ... & ... & ... & ... & ...   \\
Q2343-BX415 & X & X & ... & X & X & ... & ... & ... & X  \\
\hline
\multicolumn{10}{c}{Narrow-line AGN}\\
\hline
Q0100-BX172 & X & X & X & ... & ... & ... & X & X & ...  \\
Q0142-BX195 & X & X & X & ... & ... & X & X & X & X  \\
Q0207-BX298 & X & X & X & ... & ... & X & X & X & ...   \\
Q0821-D8 & X & ... & X & ... & ... & X & X & X & ...   \\
GOODSN-BMZ1384 & X & X & X & ... & ... & ... & ... & ... & ...   \\
Q1623-BX454 & ... & X\tablenotemark{a} & X & ... & ... & ... & X & ... & ...   \\
Q1700-MD157 & ... & X\tablenotemark{a} & X & ... & ... & ... & ... & ... & ...   \\
Q2343-BX333 & ... & X\tablenotemark{a} & X & ... & ... & X & X & ... & ...   
\enddata
\label{specsummary.table}
\tablenotetext{a}{No detection.}
\tablenotetext{b}{\Hb\ observation instead of \Ha.}
\end{deluxetable*}

\section{OSIRIS IFU Observations}
\label{osirisobs.sec}

\subsection{Observing Strategy}
\label{osiris.strategy.sec}

We obtained observations using the
Keck/OSIRIS $+$ LGSAO system during eight observing runs between
September 2007 and January 2017 with an observing strategy similar to that outlined by \citet{law07,law09,law12b}.
In brief, each half-hour exposure block consisted of two 15 minute integrations dithered along
the long axis of the rectangular IFU in an A-B sequence.  As discussed by \citet{law09}, this approach maximizes on-source integration time whilst simultaneously
obtaining local background measurements at the cost of a smaller common field of view between the pointings.
Galaxies detected after one such sequence were observed for up to
three additional hours in order to improve the detection quality.  Galaxies not detected in the initial 30-minute sequence
were generally not observed further, although in the case of Q0821-D8 longer observations were taken despite
initial non-detection due to observing window constraints on the available target sample.

Each science observation was preceded by a short ($\sim 5$ second) observation of a nearby A0V reference star
and a $\sim 60$ second observation of the LGSAO tip-tilt (TT).  These calibration observations jointly allow us to
flux calibrate the OSIRIS science observations, correct for atmospheric telluric absorption features,
provide a measure
of the LGSAO PSF, and  refine the astrometric pointing prior to observing the science target using offsets
derived from {\it HST} or ground-based imaging data.
In each case we used the appropriate narrowband order selection filter for the known wavelength of nebular line emission
based on prior slit spectroscopy.  We generally used the 50 mas lenslet scale for our observations in order to maximize the
effective angular resolution of the data, except in cases where prior {\it HST} imaging indicated that the object likely had a
$>1''$ angular size or for which prior spectroscopy suggested that the object had particularly low surface brightness.

Table \ref{targets.table} lists the individual observations obtained in our OSIRIS observing program.
In all we obtained about 33 hours of on-source integration, 30 hours of which were spent on successfully detected objects.
Our observations typically targeted \othree\ $\lambda5007$ in order to measure
the size and structure of the extended NLR, but in some cases included \Ha\ and/or \Hb\  as well 
depending on source redshift and the amount of observing time available at a given R.A..


\subsection{Data Reduction}
\label{osiris.reductions.sec}

The OSIRIS data were reduced using v4.0 of the OSIRIS data reduction 
pipeline\footnote{https://github.com/Keck-DataReductionPipelines/OsirisDRP/tree/master} in combination with custom 
IDL scripts\footnote{Available at https://github.com/drlaw1558/osiris}
as described in greater detail by \cite{law09}.  In brief, A-B exposure pairs were differenced at the raw detector level
to provide first-pass sky subtraction, and then extracted to individual data cubes using calibration reference matrices
by the OSIRIS pipeline.  Each individual cube is then sent through a second-pass sky subtraction routine that computes
the median value at each wavelength slice of the cube and subtracts it from the slice\footnote{Flux from the
science target is negligible in this process as our targets are compact, faint, and have equal positive and negative
signature in each cube because of our A-B pairwise subtraction strategy.}; this effectively removes
artifacts due to drifts in either the sky continuum or OH line intensity between exposures.  These individual exposures
are then flux calibrated and telluric corrected using observations of the A0V reference star matched to a theoretical
Vega template spectrum with the overall flux normalization
derived from matching the tip-tilt reference star spectrum against known 2MASS infrared magnitudes.
The final data cubes are then constructed by median combining the individual cubes together after application of
the astrometric solution derived from the LGSAO  header keywords, and have units of
$10^{-17}$ \cgsang\ spaxel$^{-1}$.  These cubes are masked to include only the regions of common overlap between all IFU
pointings in order to ensure that no negative artifacts from the A-B observing strategy are present in the final data cubes.

As discussed in \cite{law07} we estimate a global systematic uncertainty $\sim$ 20\% in our absolute flux calibration for
these composite data cubes
(although the relative flux ratios within
a given spectrum are much more precisely constrained) primarily caused by the uncertainty in
the structure of the LGSAO PSF, which can vary from on-axis TT star observations to
off-axis science observations, and on the timescale of minutes or less even in a fixed field.
These uncertainties are comparable to the flux calibration uncertainty of the MOSFIRE data as well, for which correction factors
for slit losses range from $\sim 1.5 - 2$ depending on the observational seeing and intrinsic morphology of the source
\citep[see discussion by][]{steidel14, strom17}.

Special processing was required for \othree\ observations of the galaxy Q0100-BX172, for which data obtained in 
September 2008 vs January 2017 differed by 40 \kms\ in their barycentric wavelength correction, requiring that
the barycentric correction be applied prior to combining individual dithered frames into a composite data cube (instead of after construction of the final cube.\footnote{Other objects observed in different seasons
has radial velocity solutions that differed by $< 10$ \kms; substantially below the $\sim 40$ \kms\ $1\sigma$ width
of the instrumental line-spread function (LSF).}  Since the adjustment
was very nearly one spectral channel (36 \kms) this shift was achieved with minimal interpolation.
Additionally, the position angle of the observations changed by 180 degrees between the two epochs
due to the different geometry of the low bandwidth wavefront sensor when OSIRIS was installed on Keck 1 (2017) vs Keck 2 (2008).
The 2008 data was therefore rotated
to the position angle of the 2017 data, and the dither offsets recomputed by hand using stacked data from the individual epochs
prior to final combination of the composite data cube.

\subsection{Spectral Fitting}
\label{osiris.fitting.sec}

The final data cubes were processed using a custom IDL routine\footnote{https://github.com/drlaw1558/osiris/osredx\_velmap.pro} that fits single gaussian components
to each spaxel by wrapping the core MPFIT algorithm \citep{mark09}.
The input data cube is first smoothed spatially at each wavelength channel with a gaussian kernel of FWHM $\sim 2$
spaxels (i.e., $\sim 0.1$ arcsec for observations using the 50 mas lenslet scale, $\sim 0.2$ arcsec for observations using the
100 mas lenslet scale) in order to better detect
faint structures; the FWHM of the smoothing kernel varies from galaxy to galaxy.
The initial spectral fitting uses a nominal error spectrum derived from the average background sky spectrum, while a second iteration of the fit
incorporates an empirical estimate of the covariance introduced by our spatial smoothing to ensure that the residual 
(data-model) spectra are consistent with
the covariant error spectrum.
In each case the width of the gaussian component is bounded to be between the instrumental spectral line spread function (LSF)
and 1000 \kms, the velocity to be within 1000 \kms\ of the redshift estimate provided by prior LRIS and/or MOSFIRE spectroscopy,
and the flux is required to be positive.
For \othree\ we tie the velocity and velocity dispersion of the $\lambda$ 5007 and 4959 components, and fix
the total flux ratio to 3.0.  
Uncertainties in the derived values of flux are set to 20\% (since the systematic uncertainty from the LGSAO flux calibration dominates over the statistical uncertainty
in the fits), while the uncertainties in the derived velocity and velocity dispersion values are 
based on the formal errors returned by the MPFIT routine.

\begin{figure}[p!]
\plotone{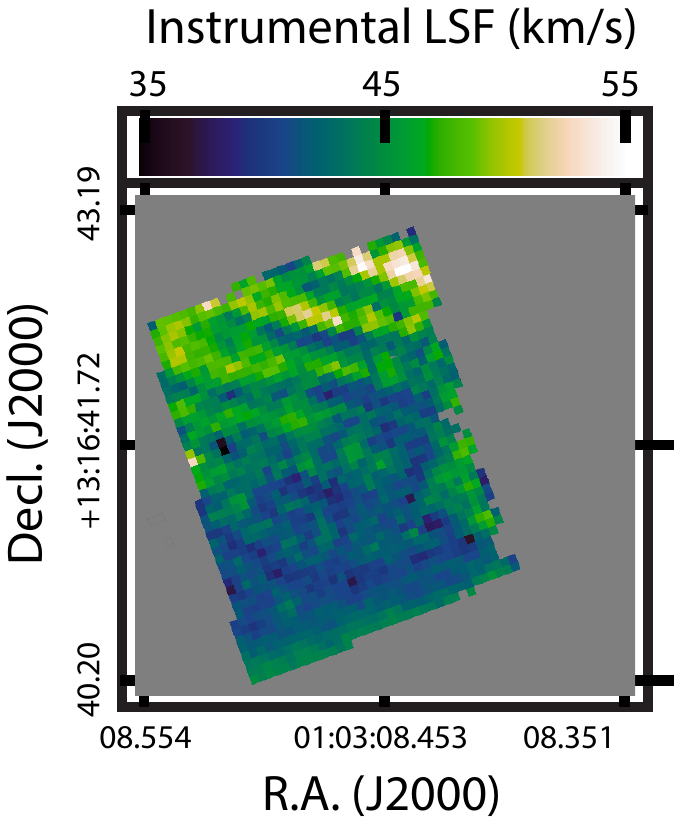}
\caption{Spectral $1\sigma$ line spread function (LSF) variation across the reconstructed data cube for Q0100-BX172 as derived
from measurements of isolated OH skylines; spatial structure represents variations
in the spectral resolving power across the OSIRIS lenslet grid.
}
\label{lsf.fig}
\end{figure}

The astrophysical $1\sigma$ velocity dispersion in each spaxel is computed as the square root of the quadrature difference between the measured width
of the emission line and the instrumental LSF.  We note that the LSF is itself spatially variable across the OSIRIS field of view due to
differences in the resolving power across the lenslet grid.  We estimate the LSF at each spaxel in our final composite data cubes by 
constructing a similar cube for which none of the exposures have been sky subtracted, and fitting single-component
gaussian models to each of a set of OH skylines that are known \citep[][]{rous00} to be relatively bright and isolated.\footnote{There are typically
1-4 such lines per data cube for the OSIRIS narrowband
filters, which is sufficient to obtain an average estimate of the LSF but insufficient to reliably model any wavelength dependence over the spectral bandpass.}
As illustrated by Figure \ref{lsf.fig}, we find that the $1\sigma$ LSF can vary over the field of view by up to 25\% (in the case of Q0100-BX172, from $\sim 40-50$ \kms).

\begin{figure*}[p!]
\epsscale{1.05}
\plotone{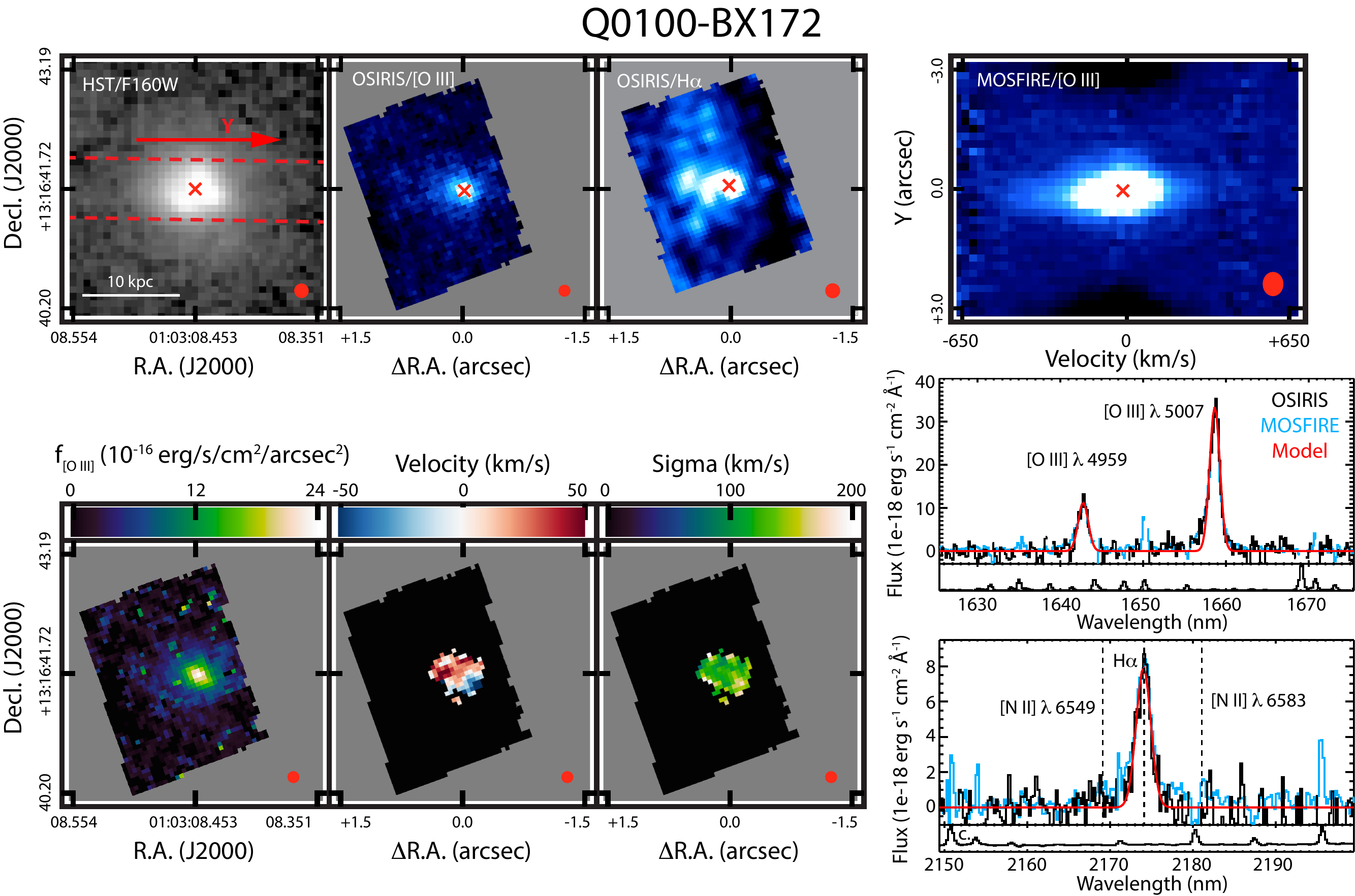}
\caption{HST/WFC3 F160W imaging, Keck/OSIRIS, and Keck/MOSFIRE \othree\ spectroscopy for Q0100-BX172.  The top panels show the F160W stellar continuum
morphology (rest-frame $\sim 4600$\AA, logarithmic color stretch), channel-collapsed OSIRIS \othree\ and \Ha\ flux maps (linear stretch), and two-dimensional MOSFIRE \othree\ spectrogram (linear stretch) in which the spectral dimension is shown in units of velocity with respect to the systemic rest frame.  Note that all panels (except the MOSFIRE spectrogram) 
have the same field of view.
The dashed lines in the first panel indicate the orientation of the MOSFIRE slit and its along-slit spatial (Y) orientation, in all panels
the red X indicates the center of the source and the red circle indicates the FWHM of the observational PSF (and spectral LSF for MOSFIRE).
The bottom left panels show the \othree\ emission line surface brightness, relative velocity, and velocity dispersion (after quadrature subtraction of the instrumental
profile) in each spaxel for the \othree\ data cube using a single component gaussian fit to the spectra.  The bottom right panels plot the integrated OSIRIS \othree\ and \Ha\
spectra of the source (black line), along with the corresponding MOSFIRE spectra (blue line), and the best-fit single component gaussian models
(red line).  Vertical dotted lines indicate the nominal wavelengths for \ntwo\ emission.  Note that the MOSFIRE spectra have been normalized to match the total line strength observed with OSIRIS.  The bottom section of the spectrum panels
illustrate the error spectrum that is dominated by the bright OH skyline emission- data in the science spectra have been masked at the wavelengths of strong OH emission in order to suppress sky subtraction residuals.
}
\label{bx172.fig}
\end{figure*}

\begin{figure*}[p!]
\epsscale{1.05}
\plotone{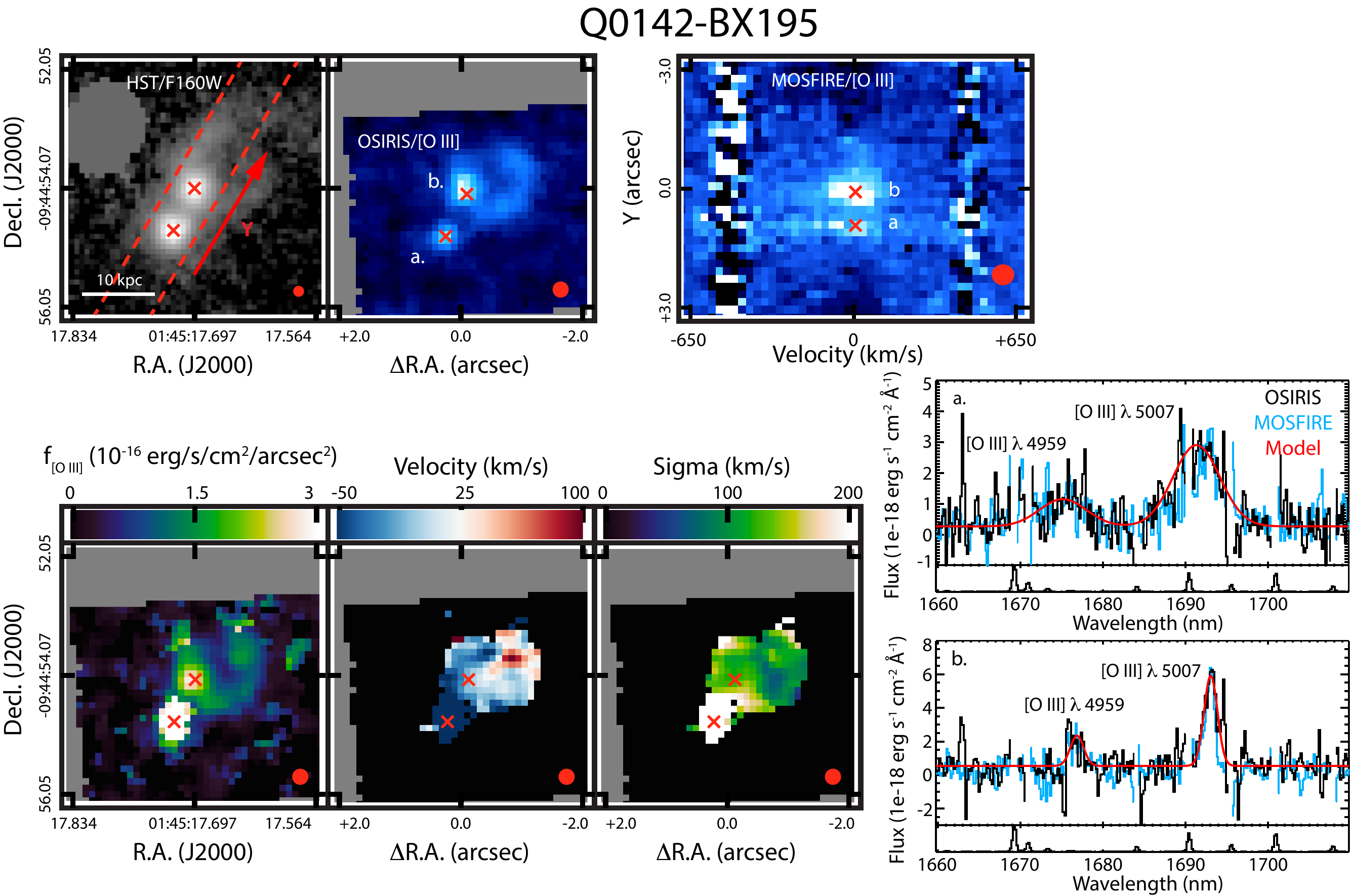}
\caption{As Figure \ref{bx172.fig}, but for Q0142-BX195.  The grey region in the {\it HST}/F160W image represents a defect on the WFC3 detector, and 
red Xs denote the location of the two nuclei (labelled [a] and [b]) in each panel.
Lower right panels show the \othree\ spectra of the two components, indicating that the [a] component is substantially broader than the [b] component.  
}
\label{bx195.fig}
\end{figure*}

\begin{figure*}[p!]
\epsscale{1.05}
\plotone{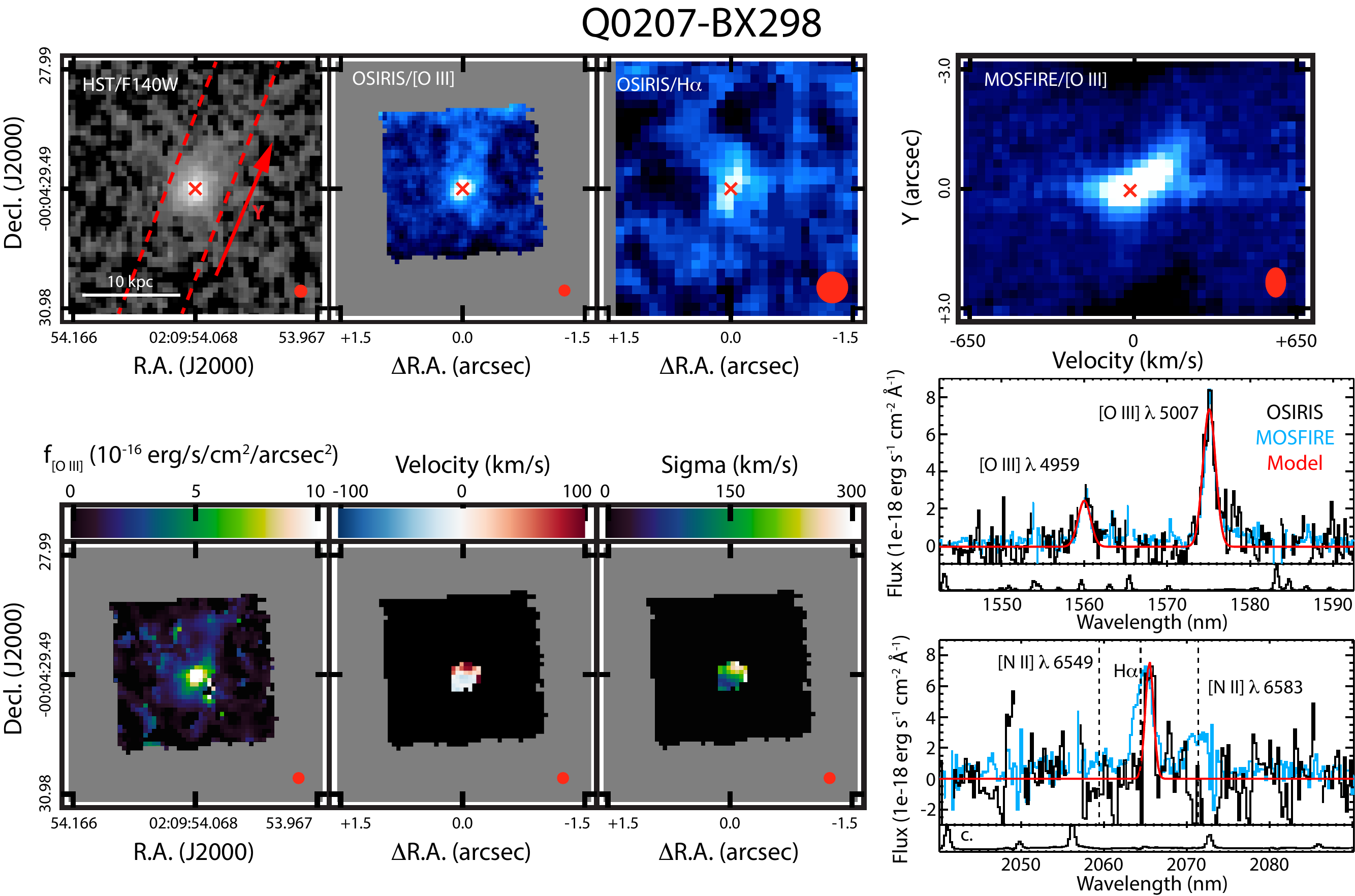}
\caption{As Figure \ref{bx172.fig}, but for Q0207-BX298.}
\label{bx298.fig}
\end{figure*}

\begin{figure*}[p!]
\epsscale{1.05}
\plotone{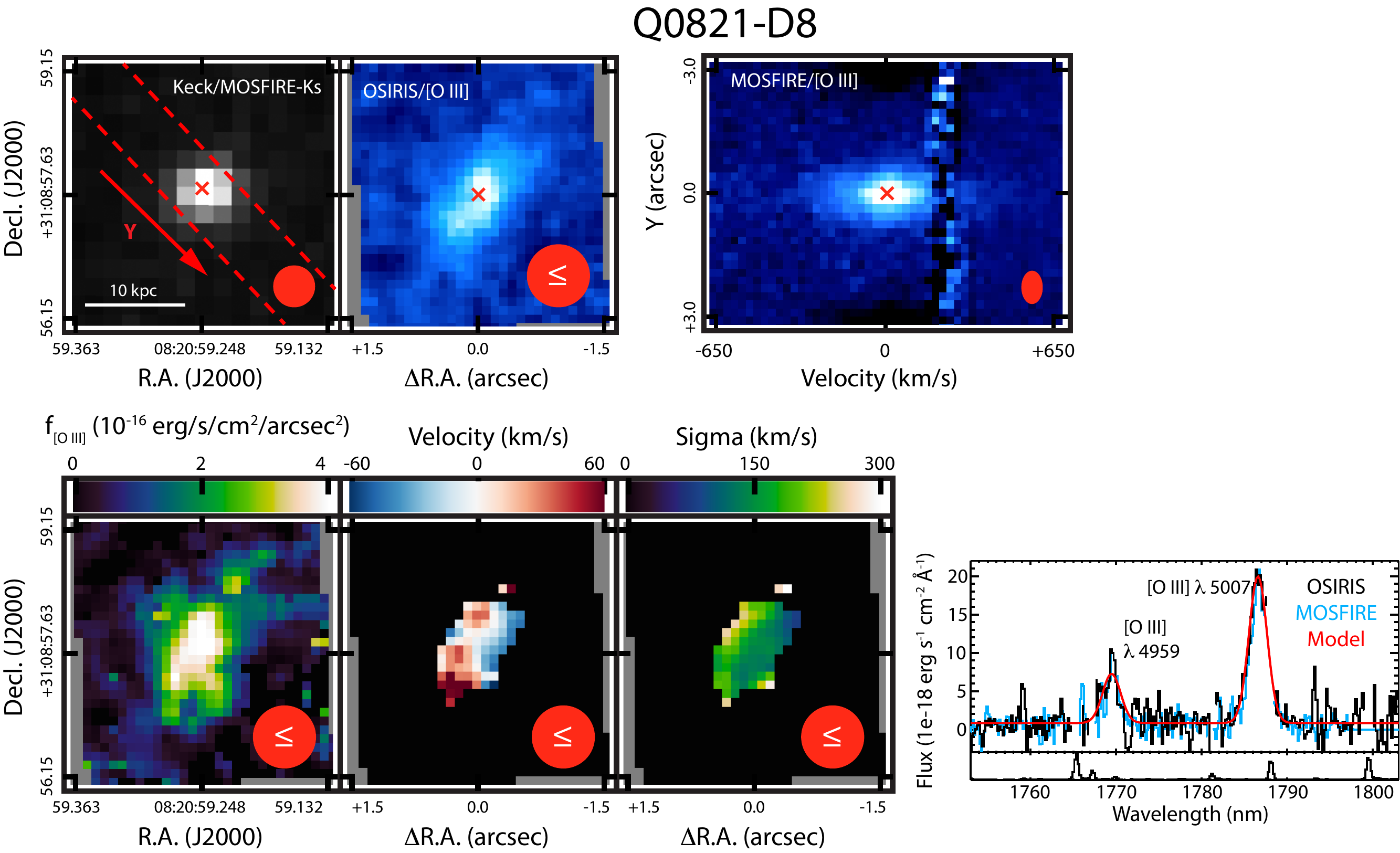}
\caption{As Figure \ref{bx172.fig}, but for Q0821-D8.  The observational PSF is shown as an upper limit corresponding to the $H$-band seeing since it
was not possible to estimate the actual PSF from observations of the faint tip-tilt source.}
\label{d8.fig}
\end{figure*}

\begin{figure*}[p!]
\epsscale{1.05}
\plotone{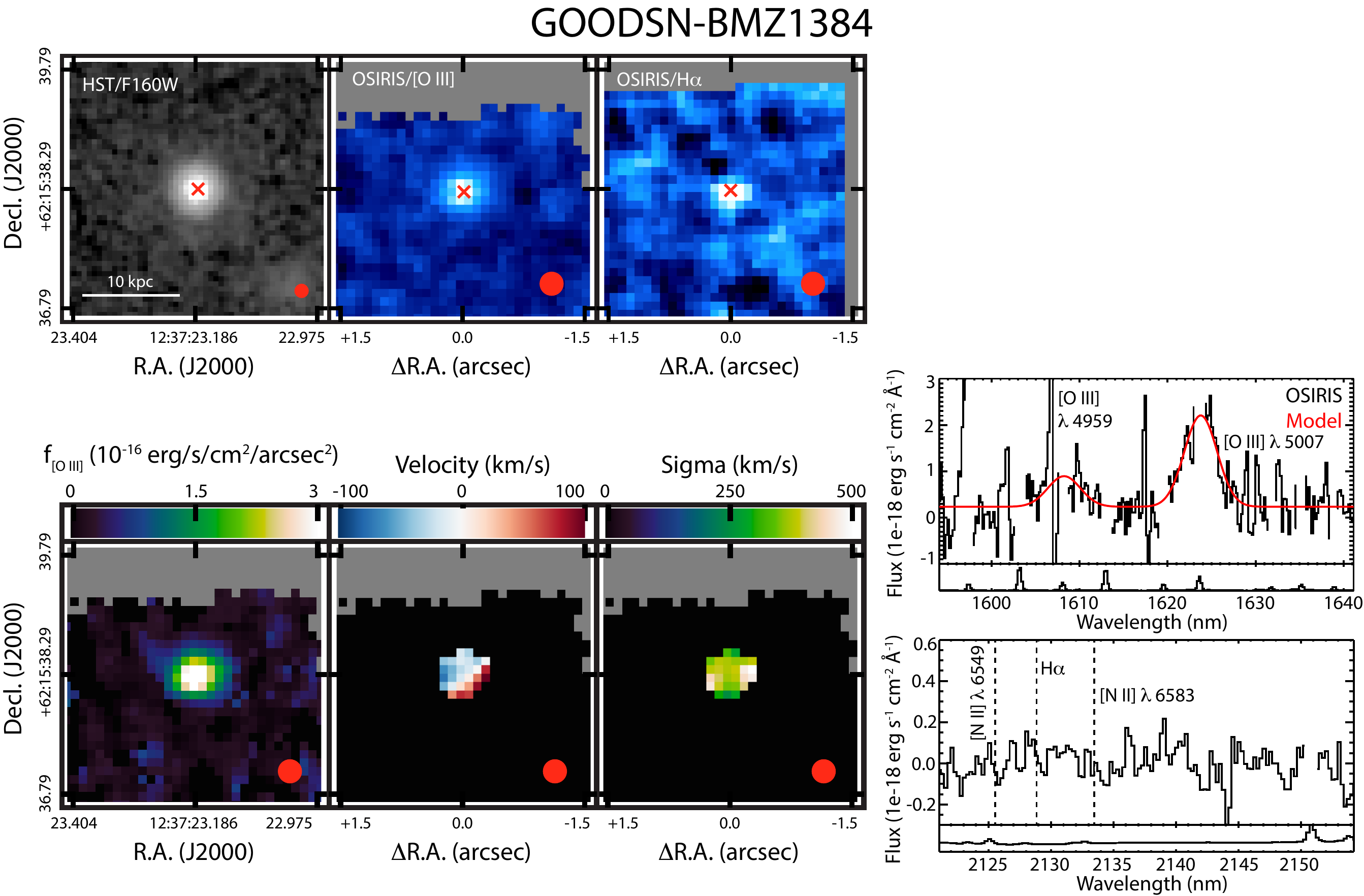}
\caption{As Figure \ref{bx172.fig}, but for GOODSN-BMZ1384.  Note that the apparent velocity gradient in the middle panel is not significant as the \othree\ emission is morphologically consistent
with a point source.}
\label{bmz1384.fig}
\end{figure*}

\begin{figure*}[p!]
\epsscale{1.05}
\plotone{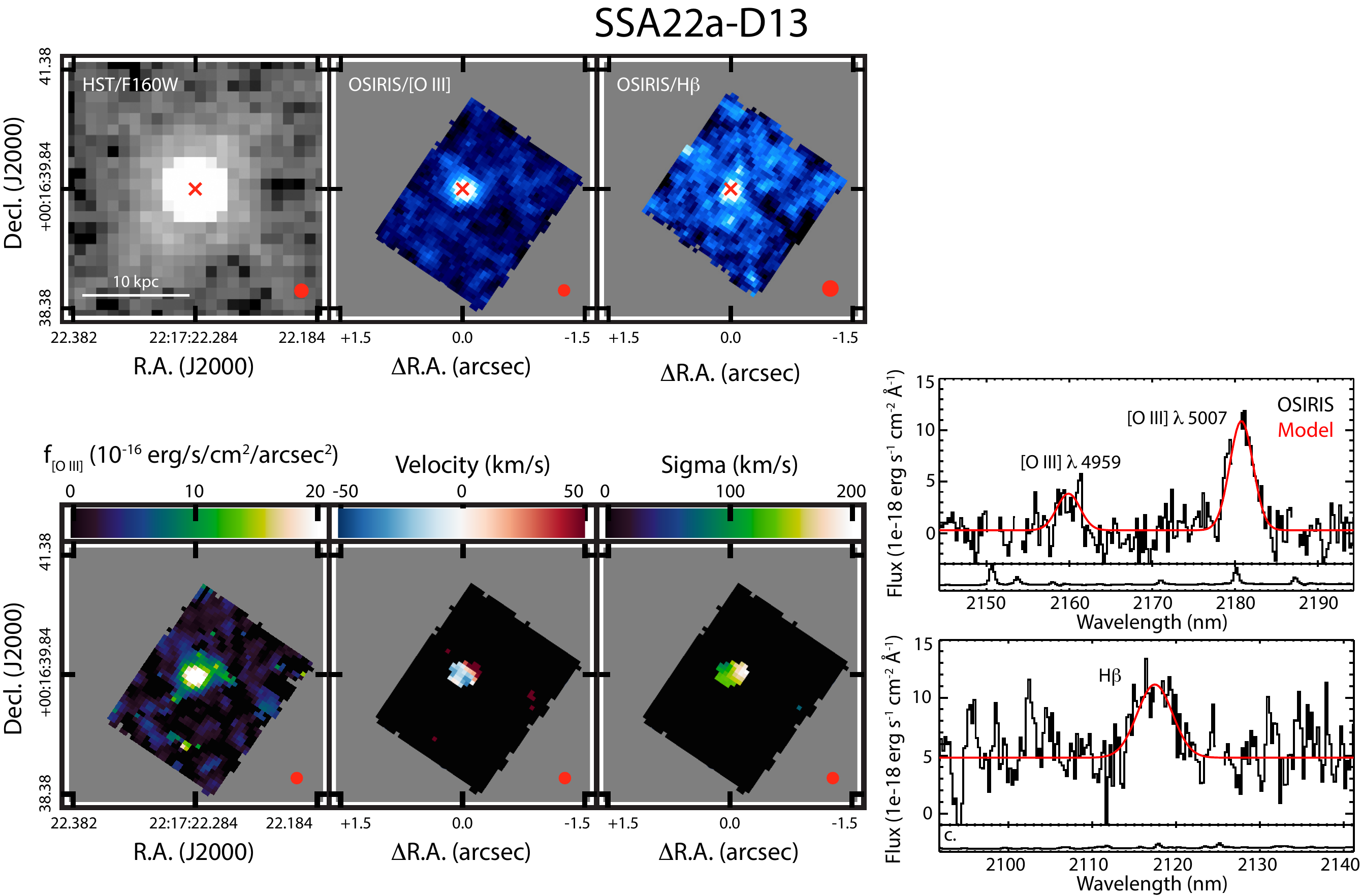}
\caption{As Figure \ref{bx172.fig} but for SSA22a-D13.}
\label{d13.fig}
\end{figure*}

\begin{figure*}[p!]
\epsscale{1.05}
\plotone{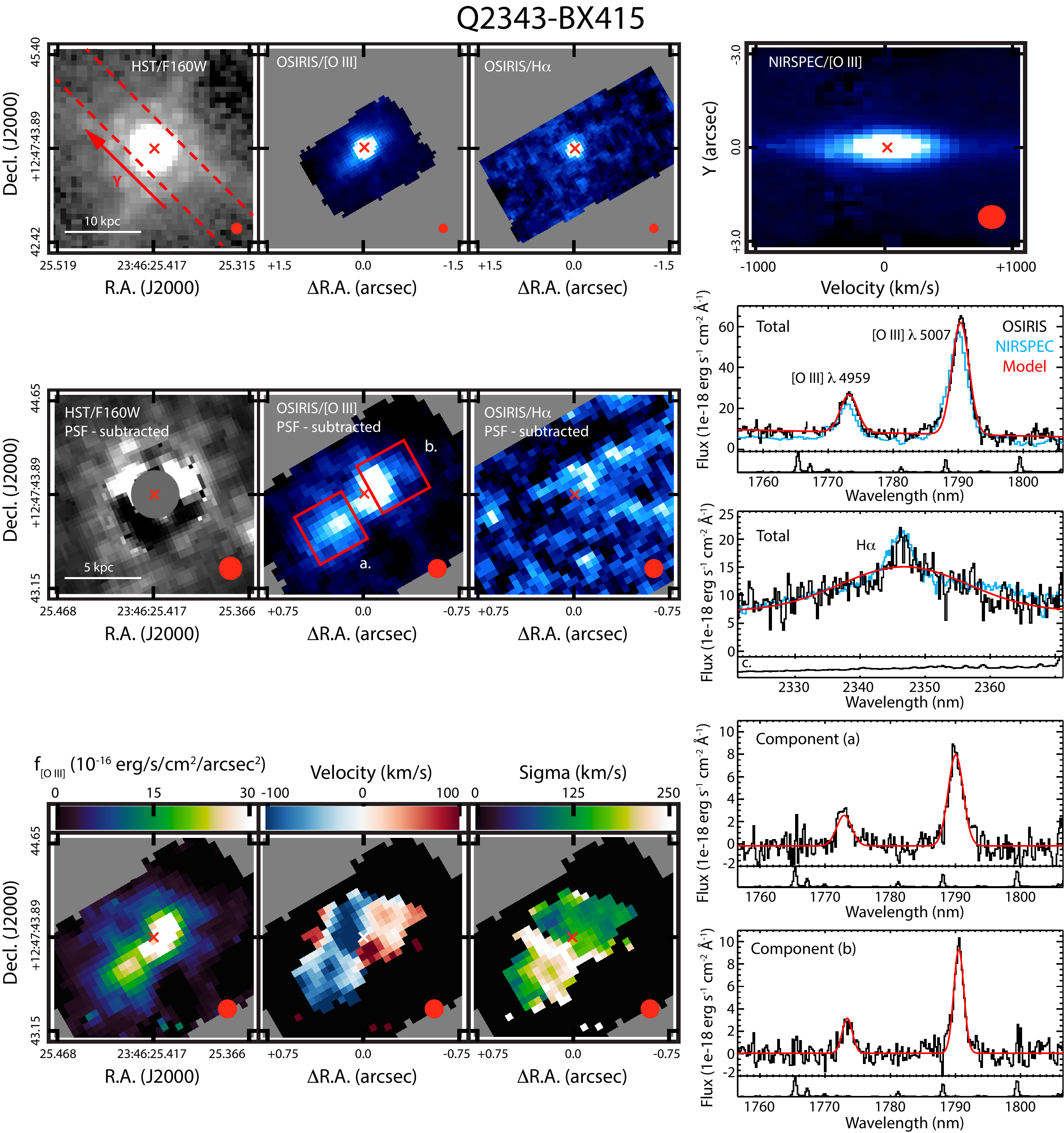}
\caption{As Figure \ref{bx172.fig} but for Q2343-BX415.  The top row of panels illustrate the observed HST continuum morphology, OSIRIS \othree\ and \Ha\ flux maps, and two-dimensional
Keck/NIRSPEC spectrogram.  Since these are dominated by a central point source the middle left row of panels shows the residual structure of the continuum, \othree\, and \Ha\ flux maps
after subtraction of the central point source.  The bottom left panels show the \othree\ emission line surface brightness, relative velocity, and velocity dispersion
of the PSF-subtracted \othree\ data cube.  Spectra in the middle-right panels show the source-integrated \othree\ and \Ha\ emission line profiles for the OSIRIS and NIRSPEC observations
compared to the fitted  model, while the spectra in the bottom-right panels show the PSF-subtracted \othree\ residual spectra of components [a] and [b] (marked as red boxes in the PSF-subtracted
\othree\ line map panel).}
\label{bx415.fig}
\end{figure*}

In Figures \ref{bx172.fig} - \ref{bx415.fig} we show the kinematic maps derived from this fitting technique for each of the seven galaxies for which our OSIRIS observations 
detected statistically significant \othree\  or \Ha\ emission, and include for comparison the {\it HST}-based continuum image and two-dimensional long-slit MOSFIRE spectrogram
(where available).
We show the \othree\ emission line morphology two different ways: first with a simple median collapse of the data cube
across the wavelength channels near peak emission (top row), and second with the emission line surface brightness resulting from
spaxel-by-spaxel fits to the spectra in the data cube (middle row).  The former best illustrates bright narrow emission features, while the latter
is a more robust method that illustrates the relative amount of total flux in both broad and narrow emission components.
Neither \othree\ flux map has a quality cut applied to the resulting data (requiring only a positive integral of the gaussian function within the spectral window),
while for the line-of-sight velocity and velocity dispersion maps we mask out all spaxels with signal-to-noise ratio (SNR) $<$ 4-5
in order to minimize confusion from spurious features.

Similarly, we use this SNR mask to optimize the quality of our extracted one-dimensional spectra
shown in the lower panels of Figures \ref{bx172.fig} - \ref{bx415.fig}.  Starting with the unmasked spaxels for which velocity information is reported in the middle row of these plots
we grow the mask by 4 spaxels in every direction and sum the spectra of all spaxels included in this new mask region.  This growth radius for the extraction region
was determined empirically where the extraction region is sufficiently large to encompass the majority of flux from the galaxy (i.e., the curve of growth of flux with increasing radius
has plateaued) yet sufficiently small that it does not include additional noise from too many spaxels that contribute negligible galaxy flux.  This dynamic approach to spectral extraction is
able to trivially accommodate any irregularities in the source geometry, although for Q0142-BX195 we instead use simple extraction boxes designed to separate the components of
the broad and narrow-lined nuclei (see \S \ref{bx195.sec}).  In addition to the composite OSIRIS spectrum we also overplot the MOSFIRE extracted one-dimensional spectra
in the lower panels of Figures \ref{bx172.fig} - \ref{bx415.fig}, applying an arbitrary scaling for display purposes to the MOSFIRE spectra such that they agree in overall normalization
with the OSIRIS spectra (i.e., we remove any systematics due to relative calibration uncertainties between the two instruments so that we can instead compare the observed shapes
of the line profiles more easily).


We tabulate the corresponding source-integrated \othree\ velocity dispersions $\sigma_{\rm tot}$
(i.e., including potential contributions from both unresolved kinematics and any resolved line-of-sight velocity gradients),
fluxes, and total luminosities for our adopted cosmology in Tables \ref{results.table} and \ref{mosfire.table}
for the OSIRIS and MOSFIRE observations respectively.  As discussed in \S \ref{sed.sec},
we use the extinctions derived from our stellar population modeling to dust-correct the total luminosity for the Type II AGN in our sample.
In addition, for the objects with MOSFIRE spectra we also tabulate the values for the source-integrated \ntwo/\Ha\ and \othree/\Hb\ diagnostic
line ratios, noting that most exhibit the characteristic line ratios expected for objects whose gas is predominantly photoionized by the hard spectrum of the central AGN
(see discussion in \S \ref{ratios.sec}).

\begin{deluxetable*}{lccccccc}
\tablecolumns{8}
\tablewidth{0pc}
\tabletypesize{\scriptsize}
\tablecaption{OSIRIS Spectral Fitting Results}
\tablehead{
\colhead{Name} & \colhead{$F_{\othree}$} & \colhead{$L_{\othree}$\tablenotemark{b}} & \colhead{PSF FWHM} & \colhead{$r_{\rm e}$\tablenotemark{c}} & \colhead{$r_{\rm c}$\tablenotemark{d}} & \colhead{$r_{\rm iso}$\tablenotemark{e}} & \colhead{$\sigma$} \\
  & \colhead{(10$^{-17}$ erg s$^{-1}$ cm$^{-2}$)} & ($10^{42}$ erg s$^{-1}$) & \colhead{(arcsec)} & \colhead{(arcsec)} & \colhead{(kpc)} & \colhead{(kpc)} & \colhead{(km s$^{-1}$)}}
\startdata
\multicolumn{8}{c}{Broad-line QSO}\\
\hline
SSA22a-D13 & $37 \pm 7$ & $40$ & 0.15 & 0.13 & 0.8 & 5 & $192 \pm 13$  \\
Q2343-BX415 & $140 \pm 30$ & $80$ & 0.15 & 0.24 & 1.5 & 17 &  $221 \pm 2$\\
\hline
\multicolumn{8}{c}{Narrow-line AGN}\\
\hline
Q0100-BX172  & $51 \pm 10$ &  $38 \pm 8$ & 0.15 & 0.24 & 1.3 & 13 & $110\pm3$  \\
Q0142-BX195A\tablenotemark{a} & $19 \pm 4$ & $17 \pm 3$ & 0.28 & $<0.11$ & $<0.9$ & $<6$ &  $500\pm30$  \\
Q0142-BX195B\tablenotemark{a} & $11 \pm 2$ & $10 \pm 2$ & 0.28 & 0.54 & 3.2 & 10 & $143 \pm 7$  \\
Q0207-BX298 & $14 \pm 3$ & $6 \pm 1$ & 0.15 & $<0.06$ & $<0.5$ & $<4$ & $145 \pm 8$  \\
Q0821-D8 & $51 \pm 10$ & $55 \pm 11$ & $\leq 0.7$ &  ... & ... & ... & $178 \pm 7$ \\
GOODSN-BMZ1384 &  $9 \pm 2$ &  $4 \pm 1$ & 0.30 & $<0.12$ & $<1.0$ & $<4$ & $340 \pm 30$  
\enddata
\tablenotetext{a}{Results are given individually for components A/B and omit the extended tidal features.}
\tablenotetext{b}{\othree\ luminosity; values for narrow-line AGN have been dust-corrected following Table \ref{sed.table}.}
\tablenotetext{c}{Effective radius of the Sersic profile fit.}
\tablenotetext{d}{Circularized effective radius $r_{\rm c} = r_e \sqrt{q}$.}
\tablenotetext{e}{Isophotal radius from extrapolating the Sersic profile fit.}
\label{results.table}
\end{deluxetable*}

\begin{deluxetable*}{lcccccccc}
\tablecolumns{9}
\tablewidth{0pc}
\tabletypesize{\scriptsize}
\tablecaption{Keck/MOSFIRE and NIRSPEC Spectral Fitting Results}
\tablehead{
\colhead{Name} &  \colhead{log $\frac{\ntwo}{\Ha}$} & \colhead{log $\frac{\othree}{\Hb}$} & \colhead{$F_{\othree}$\tablenotemark{b}} & \colhead{$L_{\othree}$\tablenotemark{c}} & \colhead{PSF FWHM} & \colhead{$d_{\rm iso}$\tablenotemark{e}} & \colhead{$r_{\rm iso}$\tablenotemark{f}} & \colhead{$\sigma$} \\
  &  & & \colhead{(10$^{-17}$ erg s$^{-1}$ cm$^{-2}$)} & ($10^{42}$ erg s$^{-1}$) & \colhead{(arcsec)} & \colhead{(arcsec)} & \colhead{(kpc)} & \colhead{(km s$^{-1}$)}
}
\startdata
\multicolumn{9}{c}{Broad-line QSOs}\\
\hline
Q0100-BX160 & ... & ... & $< 0.6$ & $< 0.2$ & 0.64 & ... & ... & ... \\
Q0100-BX164 & -1.2 & ... & $<0.6$ & $<0.2$ & 0.63 & ... & ... & ...\\
Q2343-BX415 &   ... & ... & $120 \pm 20$ & $66$ & 0.72 & 2.3 & 9.0 & $240 \pm 6$\\
\hline
\multicolumn{9}{c}{Narrow-line AGN}\\
\hline
Q0100-BX172  & $< $-1.0 & 0.97 & $64 \pm 13$ & $47 \pm 10$ & 0.62 & 2.5 & 10.3 & $114 \pm 1$ \\
Q0142-BX195A\tablenotemark{a} & -0.49 & 0.94  & $12 \pm 2$ & $11 \pm 2$ & 0.49 & 1.08 & 4.0 & $451 \pm 5$ \\
Q0142-BX195B\tablenotemark{a} & -0.49 & 0.94 &  $10 \pm 2$ & $9 \pm 2$ & 0.49 & 2.16 & 8.8 & $92 \pm 1$ \\
Q0207-BX298 &   -0.42 & 0.98 & $21 \pm 5$ & $10 \pm 2$ & 0.77 & 1.6 & 6.1 & $149 \pm 1$ \\
Q0821-D8 &   -0.36 & 0.93  &$22 \pm 4$ & $24 \pm 4$ & 0.81 & 2.2 & 8.3 & $159 \pm 2$ \\
GOODSN-BMZ1384\tablenotemark{d}   & ... & ...  & ... & 9 & ... & ... & ... & 560 \\
Q1623-BX454 & ... & 0.38 & $3 \pm 1$ & $1.5 \pm 0.5$ & 0.75 & 0.9 & $<2.3$ & $225 \pm 18$ \\
Q2343-BX333 &  ... & 0.94 & $3 \pm 1$ & $2 \pm 0.6$ & 0.57 & 0.4 & $<2.0$ & $227 \pm 13$ 
\enddata
\label{mosfire.table}
\tablenotetext{a}{Results are given individually for components A/B and omit the extended tidal features.}
\tablenotetext{b}{Total \othree\ flux assuming a factor of two slit loss correction.}
\tablenotetext{c}{\othree\ luminosity; values for narrow-line AGN have been dust-corrected following Table \ref{sed.table}.}
\tablenotetext{d}{Values taken from \citet{leung17}.}
\tablenotetext{e}{Isophotal diameter in angular units.}
\tablenotetext{f}{Isophotal radius in physical units.}
\end{deluxetable*}

\subsection{Characteristic Depth}
\label{osiris.dq.sec}

The OSIRIS instrument has evolved considerably over the 10-year baseline covered by our observations, with
upgrades in 2012 and 2016 respectively to the grating \cite[see, e.g.,][]{mieda14} and the detector \cite[replacing a
Hawaii-2 with a Hawaii-2RG detector array; see][]{boehle16}.  These upgrades together have substantially improved
the sensitivity of OSIRIS; based on observations of Q2343-BX415 spanning the time interval Sep 2007 - October 2016 we 
estimate a 1$\sigma$ senstivity of $5$ and $3 \times 10^{-19}$ \cgsang\ spaxel$^{-1}$ respectively for a
single 15-minute exposure in good conditions in 2007 vs 2016.

Indeed, all five of the sources that we observed with OSIRIS but did not detect were observed early in our
observational program when the instrument sensitivity was lower and focused on \Ha\ rather than \othree\ emission.
This 60\% success rate is comparable to that of our observations of the star forming parent 
population of $z \sim 2$ KBSS galaxies \citep[54\%,][]{law09}.
As indicated by our more recent MOSFIRE observations however, these five sources are genuinely faint
and would not have been detected in \othree\ with OSIRIS even after the various upgrades.


The final depth of the stacked data cubes 
illustrated in Figures \ref{bx172.fig} - \ref{bx415.fig} varies from galaxy to galaxy depending on the total exposure time, 
thermal $+$ OH skyline background, lenslet scale, and applied spatial smoothing kernel.
Typically, we estimate the $3\sigma$ sensitivity within a $0.2 \times 0.2$ arcsec box (i.e., corresponding roughly to a spatial
resolution element) to be $2-4 \times 10^{-17}$ \cgsangas\ ($\sim 3 \times 10^{-16}$ \cgs\ arcsec$^{-2}$ integrated over a typical narrow-line profile) for the \othree\ observations
\citep[i.e., similar to][]{nesvadba17}, 
ranging up to 
$10^{-16}$ \cgsangas\ 
for \Ha\ observations of Q2343-BX415 at $2.34$ \micron\ where the thermal background from the optical system
is significant.  Our observations of Q0142-BX195 in particular reach lower surface brightness sensitivities than the 
other observations because of the larger 100 mas spaxels and the greater degree of spatial smoothing.

As discussed by \citet{steidel14}, our MOSFIRE observations reach considerably greater depths due to a combination of the greater throughput of the MOSFIRE optical system
(which does not suffer from AO system losses) and the larger pixel sizes (0.18'' per pixel for MOSFIRE vs 0.05'' per spaxel for OSIRIS).  
We estimate that the $1\sigma$ rms of the 2d $H$-band MOSFIRE spectrograms for our galaxies in regions away from bright OH features is roughly
$5 \times 10^{-20}$ erg s$^{-1}$ cm$^{-2}$ \AA$^{-1}$ pixel$^{-1}$, corresponding to a $3\sigma$ surface brightness sensitivity
of $5 \times 10^{-18}$ erg s$^{-1}$ cm$^{-2}$ arcsec$^{-2}$ for the 0.7 arcsec slit and a typical emission line with FWHM 250 km s$^{-1}$ ($\sim 14$ \AA).
Integrating along the slit for a typical distance of $\sim 1.5$ arcsec ($\sim 9$ pixels) we note that this corresponds to a $5\sigma$
integrated source sensitivity of $3 \times 10^{-18}$ erg s$^{-1}$ cm$^{-2}$ ($6 \times 10^{-18}$ erg s$^{-1}$ cm$^{-2}$ with a factor of 2 slit-loss correction), 
consistent with values estimated previously for the broader KBSS survey \citep{steidel14,strom17}.

\subsection{Observational PSF}
\label{osiris.psf.sec}

In order to determine the intrinsic size of our galaxies it is essential to know
the effective width of the LGSAO PSF in each OSIRIS observation.  We therefore 
produce a reference data cube combining all observations of the tip-tilt star for each target, collapse it over all wavelengths, and smooth it
in an identical manner to the science data to obtain an estimate of the PSF FWHM.
Values for each of our targets are tabulated in Table \ref{results.table} and illustrated as red filled circles in Figures 
\ref{bx172.fig} - \ref{bx415.fig}.
Of our seven well-detected sources, two are consistent with being spatially unresolved (Q0207-BX298, GOODSN-BMZ1384),
two have symmetric and measurable spatial extent (Q0100-BX172, SSA22a-D13), two have complex spatial structure (Q0142-BX195, Q2343-BX415), 
and one cannot be reliably assessed (Q0821-D8).

\begin{figure}[p!]
\epsscale{1.2}
\plotone{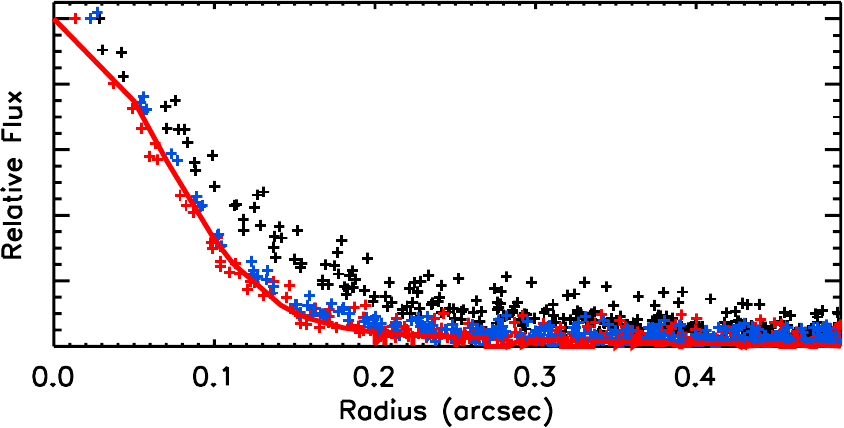}
\caption{Radial profile of Q2343-BX415 at wavelengths corresponding to \othree\ emission (black points) and collapsed over continuum-dominated
wavelengths (blue points).  The continuum radial profile is very similar to the on-axis observations of the tip-tilt star that we typically use as the reference
PSF for our OSIRIS observations (red points).  The solid red line represents a two-component gaussian model fit to the reference PSF that has a core FWHM as measured from the observations and a halo component with FWHM comparable to the observational seeing.
}
\label{radprof415.fig}
\end{figure}

We note that the utility of the tip-tilt star observations as a PSF reference is necessarily limited and approximate
because 1) The off-axis PSF delivered by the LGSAO system during science observations is known to differ from the on-axis
PSF during observations of the tip-tilt star, and 2) The PSF can vary rapidly with time as the atmospheric turbulence profile
evolves.  However, our observations of Q2343-BX415 give us confidence that our on-axis model is a reasonable approximation
for our present purposes: Unlike typical $z\sim 2$ galaxies, Q2343-BX415 is sufficiently bright
that our OSIRIS observations detect continuum emission from the 
compact region around the central QSO.  The data cube collapsed over continuum wavelengths away from
potentially-extended emission line features therefore provides a real-time way to monitor the true PSF in the data cube.
As we show in Figure \ref{radprof415.fig} (blue points), the radial profile of this realtime continuum PSF is nearly 
indistinguishable from the tip-tilt derived PSF.\footnote{This similarity is aided by the spatial smoothing that we have applied
to the data cubes, since the smoothing kernel represents a significant contribution to the total effective PSF.}

In comparison, our MOSFIRE observations are seeing-limited but obtained in excellent observing conditions and well-characterized through cotemporal observations
of PSF reference objects with the MOSFIRE slitmask.  Best-fit values for the PSF FWHM in each set of observations are given in Table \ref{mosfire.table}, and typically
range from $\sim 0.5$ to $0.75$ arcseconds.

\section{Results for Individual AGN}
\label{results.sec}

%

\subsection{Q0100-BX160}

Q0100-BX160 is a faint (${\cal R_{\rm AB}} \sim 24$) broad-lined QSO with strong rest-UV spectral features,
including pronounced \nfive\ $\lambda 1240$ emission on the red wing of the \lya\ profile.
We do not detect statistically significant \othree\ emission from Q0100-BX160 in our MOSFIRE spectroscopy, but find spatially unresolved \Ha\ emission with
a characteristic velocity dispersion $\sigma = 620$ \kms.  Although this object was not detected in our OSIRIS \Ha\ observations of the source,
this is perhaps unsurprising since broad, faint \Ha\ emission features have proven difficult to detect with the typical limiting sensitivity of OSIRIS.
{\it HST} imaging of this source (Figure \ref{hst_all.fig}) suggests an isolated point source with no indication of extended or multi-component
morphology.  Indeed, 
following \citet{law12a} we use GALFIT \citep{peng02,peng10} to construct a morphological model
of the system and find that the rest-optical continuum is consistent with an unresolved point source.

\subsection{Q0100-BX164}

Similarly to Q0100-BX160, Q0100-BX164 is a broad-lined QSO undetected in both our OSIRIS observations and in MOSFIRE $H$-band observations, placing a limit of 
$L_{\othree} < 2 \times 10^{41}$ erg s$^{-1}$ on the total \othree\ luminosity.  It is detected in our MOSFIRE $K$-band observations, exhibiting a spatially compact
region of broad \Ha\ emission with $\sigma = 1100$ \kms.  As indicated by Figure \ref{hst_all.fig} the rest-optical continuum morphology
consists of a central point source embedded within an extended galactic envelope.
Using GALFIT we find a two-component model strongly preferred in which a point source resides at the center of an extended envelope with circularized effective radius
$r_{\rm c, HST} = 3.4$ kpc.  Such a large effective radius is unusual for the $z \sim 2$ star forming galaxy sample \citep[see, e.g.,Fig 11. of][]{law12a}, suggesting that the QSO may be hosted by a particularly massive and/or extended galaxy.

Figure \ref{hst_all.fig} also shows that there is a secondary source located $\sim 2$'' to the southeast of Q0100-BX164 that is of uncertain physical association.  Ground-based photometry
suggests that such an association is unlikely as the secondary source has optical colors $U_n - G = 0.54$ and $G - {\cal R} = 1.1$ that are more consistent with lower redshift galaxy populations than with
$z \sim 2-3$ star forming galaxies.  However, the secondary source was included on the MOSFIRE $H$ and $K$-band slits and we observe both 
continuum emission at the appropriate angular separation and marginal evidence for line emission at 21662.7 \AA.
This line emission is only significant at the $3\sigma$ level, but if it can be positively identified as \Ha\ emission 
it would suggest that the secondary component has a redshift
$z = 2.2999$ (i.e., within 700 \kms\ of the systemic redshift of Q0100-BX164) and thus that Q0100-BX164 may be involved in a close-pair major merger.\footnote{Preliminary analysis of additional rest-UV spectroscopy obtained during the review of this manuscript is also inconclusive.}


\subsection{Q0100-BX172}
\label{bx172.sec}

Q0100-BX172 is part of a triplet of AGN (with BX160 and BX164), all of which lie within 30 arcsec on the sky
and within $\sim$ 1500 \kms\ in redshift.  Slit spectroscopy indicates that this galaxy contains a Type II narrow-line AGN, with detections of \lya\, \cfour\ $\lambda 1550$, \hetwo\ $\lambda 1640$, and
\hetwo\ $\lambda 4686$ in emission.

As illustrated in Figure \ref{bx172.fig}, Q0100-BX172 has an extended roughly circular envelope in HST rest-optical
imaging, with a circularized effective radius $r_{\rm c,HST} = 2.0$ kpc.
For the stellar mass of $6.3 \times 10^{10} M_{\odot}$ this radius is consistent with 
the size of typical star forming galaxies in the KBSS sample \cite[see, e.g.,][]{nagy11, law12a}.
A similar extended envelope is evident in the OSIRIS \othree\ flux map (with a GALFIT-derived $r_{\rm c} = 1.3$ kpc)
and in the MOSFIRE 2d spectrogram for which the measured FWHM of 0.80 arcsec corresponds to an effective radius of 1.8 kpc after deconvolution
of the observational PSF.
We note that
although \othree\ emission falls within the F160W bandpass, the observed line flux of 
 $5 \pm 1 \times 10^{-16}$
\cgs\  is a factor of three to four fainter than the optical continuum emission
($H_{160} = 21.93$ AB, or $\sim 2 \times 10^{-15}$ erg s$^{-1}$ cm$^{-2}$ integrated across the F160W bandpass).

Q0100-BX172 exhibits little evidence of velocity shear on scales $\gtrsim 30$ \kms.
Although the small size of the galaxy relative to the observational PSF will smear out signatures of rotation \citep[see, e.g.,][]{burkert16}, 
the slight velocity gradient apparent in Figure \ref{bx172.fig} (middle panel, middle row) is only significant at the $\sim 3\sigma$ level, and may simply represent
correlated noise in the spectral fitting near the edge of the galaxy.  
Likewise, the MOSFIRE 2d spectrogram shows no indication of velocity shear at fainter surface brightness levels in the orientation probed by the spectroscopic slit (although the slit is oriented
$\sim 45^{\circ}$ away from the apparent OSIRIS-derived kinematic major axis).

Similarly, there is no obvious structure to the velocity dispersion map with an average per-spaxel velocity dispersion of $96$ \kms\ comparable
to the source-integrated velocity dispersion of $110 \pm 3$ \kms\ (\othree) and $96 \pm 7$ \kms (\Ha).
As indicated by Figure \ref{bx172.fig}, the integrated one-dimensional spectra derived from OSIRIS and MOSFIRE
are consistent with each other; the MOSFIRE \othree\ spectrum has a velocity dispersion of $114 \pm 1$ \kms.
Both the OSIRIS and MOSFIRE spectra show evidence for
a blue wing to the \othree\ emission line profile however that is not captured by the single gaussian model.
A multi-component profile fit suggests that the central component has a velocity dispersion $\sigma = 89$ \kms with a broad wing
extending $\sim 500$ \kms blueward of the systemic redshift.

Intriguingly, the OSIRIS \Ha\ map shows clumpy structure at low surface brightness, with unresolved knots of \Ha\ emission
extending up to 10 kpc in projection away from the center of the galaxy.  The nature of these features is difficult to ascertain since each of the
clumps has total flux $2-4 \times 10^{-17}$ erg s$^{-1}$ cm$^{-2}$ and is significant at the $\sim 3-5\sigma$ level.  If genuine (and not a product of the
spatial and spectral smoothing kernels applied to the data) they may therefore correspond to small knots of star formation at rates
$\sim 10 M_{\odot}$ yr$^{-1}$.

\subsection{Q0142-BX195}
\label{bx195.sec}

Q0142-BX195 is another narrow-line system whose {\it HST} rest-optical imaging reveals it to be a complex merging system
with a double nucleus (components a. and b., separated by 0.8'', or $\sim 6$ kpc at $z=2.38$) and clear tidal features (Figure \ref{bx195.fig}).
A GALFIT decomposition of the system suggests that the two nuclei have the same integrated F160W magnitude to within $\sim 2$\%,
although component (a) is considerably more compact with $r_{\rm c,HST} \leq 0.6$ kpc (i.e., effectively unresolved) vs $r_{\rm c,HST} = 1.3$ kpc for component (b).
Assuming that the stellar mass in each component scales with the F160W continuum flux this suggests equal masses for the two components.
We note marginal evidence that the IRAC detection of this source is centered over component (a), although the $\sim 1.7$'' FWHM of the {\it Spitzer} data makes
this assessment unreliable.

\begin{figure*}[p!]
\epsscale{1.2}
\plotone{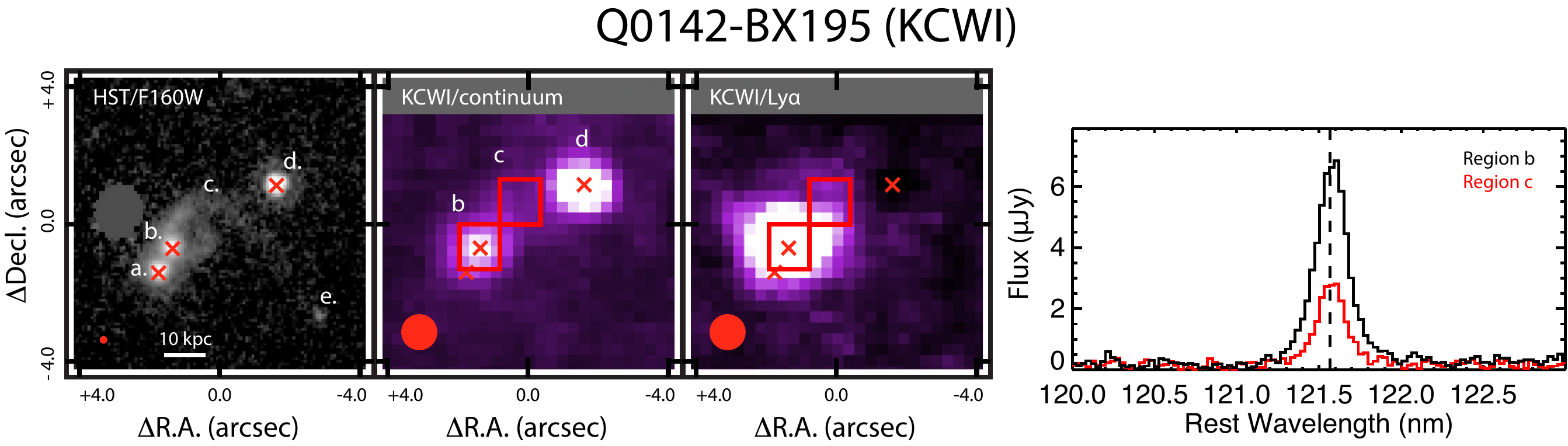}
\caption{
HST/WFC3 F160W imaging (left) and Keck/KCWI integral-field spectroscopy of Q0142-BX195
illustrating both the rest-UV continuum (middle-left) and continuum-subtracted \lya\ emission 
morphology (middle-right).
Note that the field of view shown in this figure is larger than in previous figures, and shows the
extension of the tidal features (c) down towards feature (e) which is of unknown association
with the galaxy.  Features a, b, and d are marked with red crosses in each figure, with 
feature d (a foreground star) taken to be the astrometric reference point between the HST and KCWI
frames.  Filled red circles in the lower-left corner of each panel indicate the FWHM of the observational
PSF.  The red boxes in the upper middle panel illustrate the locations at which spectra are
extracted from the IFU data cube and plotted in the right-hand panel, representing the
core of the galaxy (Region b) and the tidal feature (Region c).  The \lya\ emission feature is located at approximately the systemic redshift (dashed line)
and has velocity dispersion $\sigma = 200$ \kms.
}
\label{bx195lya.fig}
\end{figure*}

As illustrated in Figure \ref{bx195lya.fig}, with a wider field of view the {\it HST}/F160W 
tidal tail (c) extending from components (a) and (b) initially
arc toward component (d) (whose spectrum and compact nature identify it as a Milky Way foreground star) before turning towards
component (e) of uncertain physical origin.  The tidal features therefore appear to stretch at least 50 kpc in projected distance
away from the two nuclei.
Curiously, the brighter northern arc visible in the HST image is absent in \othree\ emission, while the fainter arc on the western side of the galaxy
has the brighter \othree\ counterpart.  Indeed, this western arc is consistent with having negligible stellar continuum emission since its surface brightness
(0.02 counts pixel$^{-1}$, or $24.7$ AB arcsec$^{-2}$) is comparable to the \othree\ contribution
of $10^{-16}$ \cgs\ arcsec$^{-2}$ within the F160W imaging filter bandpass.

Both nuclear components are detected in our OSIRIS \othree\ observations:
The southeastern component (a) is
broad ($\sigma = 500 \pm 30$ \kms) and offset 6.5 kpc in projection and $320 \pm 30$ \kms\ blueward 
of the northwestern component (b), which is narrower
($\sigma = 143 \pm 7$ \kms) and about half as bright as component (a).
Similarly to the F160W broadband image, GALFIT decomposition of the OSIRIS \othree\ surface brightness map indicates that component (a) is spatially unresolved
while component (b) has an effective radius $r_{\rm c} = 3.2$ kpc (although this estimate may be biased by the difficulty of deblending the tidal features).
As indicated by Figure \ref{bx195.fig} (upper right panel), both components are well-detected in our deep MOSFIRE observations
for which the slit was aligned with the morphological major axis and give $\sigma = 451 \pm 5$ and $92 \pm 1$ \kms respectively.
Although the MOSFIRE and OSIRIS values are roughly consistent with each other, the differences 
may reflect both observational uncertainty due to OH skyline residuals and intrinsic variation in the measured line profile according to small changes in the 
spatial region from which a one-dimensional spectrum is extracted.

Fortunately, our KCWI integral-field spectroscopy of the Q0142-BX195 region 
(Figure \ref{bx195lya.fig}) sheds additional light on the physical interpretation of this system since we can use component d
(a foreground star with Ca H+K absorption lines in its spectrum) as an astrometric reference point between 
the {\it HST} and KCWI observations.  Thus aligning the two observations, we note that 
both the UV continuum and UV emission-line features (\lya\ and \cfour\ $\lambda1550$) for Q0142-BX195 are centered on
component b (i.e., the narrower \othree\ source)
and to within the limitations of the $\sim 1.2$'' seeing-limited
KCWI PSF there appears to be no rest-UV counterpart to component (a).  The UV continuum flux extends slightly along the northernmost (optically bright) tidal
arm as does the \lya\ emission, for which the kinematic offset of the gas in the tidal arms may promote the escape of the resonantly scattered \lya\ photons.


Intriguingly, this means that component (a) is likely an AGN since it has an \othree\ 
velocity dispersion $ \sigma \sim 7$ times higher than typically observed for star forming galaxies,
while component (b) is likely {\it also} an AGN since it appears to be the source of the hard ionizing photons that give
rise to the rest-UV emission features that we initially used to select the system.
Perhaps most plausibly, Q0142-BX195 may therefore represent one of the highest redshift
known examples of a rare {\it double-AGN close pair major merger}
\citep[see, e.g.,][for local examples]{rosario11,comerford15,hainline16} with a 
largely gas-free tidal arm containing mostly evolved stars and a gas-rich arm in which triggered star formation 
may be occurring.
Indeed, the velocity dispersion in the western tidal feature is lower than in the central regions, with $\sigma \sim 80$ \kms (i.e., comparable to typical $z \sim 2$ star forming galaxies).
Unfortunately, the OSIRIS Hn4 narrowband bandpass of our observations is not wide enough to include \Hb, and our observations of this system in
\Ha\ were obtained in suboptimal observing conditions and
are too shallow to allow us to determine conclusively whether the diagnostic emission line ratios of this tidal feature are more consistent with photoionization
by star formation processes than the rest of the galaxy.

If we assume that the \othree\ emission from the tidal features (which has dust-corrected $L_{\othree,{\rm tidal}} = 1.5 \times 10^{43}$ erg s$^{-1}$ comparable to the \othree\ luminosity in component b)
corresponds to triggered star formation in an environment similar to to typical $z \sim 2$ star forming galaxies we can estimate the corresponding triggered star formation rate.
Taking the typical nebular line ratio of log$(\othree/\Hb) \approx 0.55$ for the KBSS sample from \citet{steidel14} and the canonical ratio $\Ha/\Hb = 2.85$ for case B recombination, we
estimate the likely corresponding \Ha\ luminosity to be $L_{\Ha} \approx 1.2 \times 10^{43}$ erg s$^{-1}$, corresponding to a star formation rate $\sim 50 \pm 25 \, M_{\odot}$ yr$^{-1}$
using the \citet{kennicutt94} relation.  Such a SFR is highly uncertain as the appropriate conversion factor depends upon the (unknown)
metallicity, dust correction factor, and source of ionizing photons in the tidal feature, but the value derived is consistent with SFR typically observed in the KBSS star forming galaxy sample \citep[e.g.,][]{erb06,steidel14}.

\subsection{Q0207-BX298}

Q0207-BX298 is  an isolated narrow-line AGN whose rest-optical continuum morphology (Figure \ref{bx298.fig}) is barely resolved in the F140W imaging data ($r_{\rm c,HST} = 0.6$ kpc ).
The OSIRIS \othree\ and \Ha\ line maps show similarly little structure and are consistent with an unresolved point source despite the visual impression in Figure \ref{bx298.fig}
(middle panel, top row) of faint diffuse emission extending to the north of the galaxy (and visible also in the HST imaging).
Given the unresolved nature of this object, the OSIRIS velocity and velocity dispersion maps necessarily show no meaningful structure with a single \othree\ component well-described
by a gaussian with $\sigma = 145 \pm 8$ \kms.


In contrast, the MOSFIRE picture of this system is dramatically different.  Fortuitously, the MOSFIRE slit was aligned reasonably well with the low surface-brightness
extension suggested by the {\it HST} and OSIRIS maps, and detects substantial \othree\ flux that has too low surface brightness to have been reliably measured in our
OSIRIS spectra.  Q0207-BX298 shows significant velocity shear along the slit (Figure \ref{bx298.fig}, upper right panel), from -45 \kms\ southeast of the galaxy core
to +135 \kms northwest of the core, corresponding to a total velocity gradient of 180 \kms\ across a distance of about 10 kpc.  This velocity shear may represent either an
outflowing wind being driven from the galaxy or (given that $\sigma = 64$ \kms far from the nucleus) intrinsic rotation of the ionized gas within the host galaxy itself.  
Despite the presence of a rotating component, the velocity dispersion of the nuclear region ($\sigma = 130$ \kms) dominates the total integrated \othree\ velocity dispersion
($\sigma = 149$ \kms) and is nearly double the value observed in ordinary star forming galaxies of comparable $M_{\ast} = 3 \times 10^{10} M_{\odot}$.
Even if the extended feature does trace rotation within the host galaxy however there is little evidence to suggest that there is significant star formation occurring within the galaxy; far
from the central AGN we find log$(\othree/\Hb) > 1.0$, suggesting that the entire system is photoionized by the hard ionizing spectrum of the central AGN.

We additionally note that in the central region of the galaxy the MOSFIRE spectrum shows an extended red wing to the \othree\ profile that reaches to $\sim +1000$ \kms\ from
the systemic redshift.  This broad component was too low surface brightness to be detected in the OSIRIS observations (although there may be some hint of its presence within the noise),
and is suggestive of a strong wind driven by the central AGN.




\subsection{Q0821-D8}

Although we obtain a robust integrated detection of \othree\ emission from the narrow-line AGN Q0821-D8 (Figure \ref{d8.fig}), the LGSAO correction on our OSIRIS observations was too poor
to determine its spatial profile reliably.  The poor LGSAO correction in this case resulted from the paucity of available guide stars for the tip-tilt correction to the wavefront; since no
stars were available we instead used a faint ($R = 18.5$) low-redshift galaxy as our tip-tilt source.  Indeed, although we were able to successfully maintain an AO lock on this object on
the low bandwidth wavefront sensor it is so faint that it was not detected in our dispersed OSIRIS reference observations and we therefore cannot measure the on-axis PSF for
our observations of Q0821-D8.  We have conservatively taken the 0.7 arcsec $H$-band seeing as an upper limit on the size of the PSF, and note that there may be 
non-circular artifacts produced by the poor AO correction.
While no HST imaging is available for this object either, deep Keck/MOSFIRE $Ks$-band seeing-limited imaging suggests that Q0821-D8 is 
consistent with an unresolved point source in the rest-optical continuum.

Despite the low-quality OSIRIS observations and lack of HST imaging, Q0821-D8 is nonetheless a valuable addition to our sample thanks to the high-quality MOSFIRE spectroscopy that
indicates that the galaxy is spatially resolved in \othree\ emission with a nearly gaussian profile and a FWHM of 0.95 arcsec.  The source-integrated MOSFIRE and OSIRIS spectra 
are also consistent with each other and indicate that the observed velocity dispersion is nearly double the typical value observed in ordinary star forming galaxies
of similar mass ($M_{\ast} \sim 10^{10} M_{\odot}$).  While the measure OSIRIS velocity dispersion of $\sigma = 178$ \kms\ is slightly larger than the MOSFIRE value
($\sigma = 159 \pm 2$ \kms) this discrepancy may be due to the slightly different regions of the galaxy traced by the IFU vs the MOSFIRE slit.



\subsection{GOODSN-BMZ1384}

As illustrated by Figure \ref{bmz1384.fig}, narrow-line AGN GOODSN-BMZ1384 (aka GN 24192) shows no discernible spatial struture
in either the rest-optical continuum, \othree\, or \Ha\ morphology.
The large \othree\ velocity dispersion $\sigma = 340 \pm 30$ \kms\ suggests however that the ionized gas predominantly traces regions photoionized by the central AGN.
We note that our OSIRIS observations are broadly consistent with prior MOSFIRE observations of the galaxy by \citet[][their object 10421]{leung17} who derive a velocity
dispersion of $\sigma = 560$ \kms\ and note no evidence of a spatially offset outflowing component.  The higher velocity dispersion of the MOSFIRE observations is driven by the presence
of a faint blue wing reaching to $\sim -1000$ \kms\ that was not obvious at the shallower depth of the OSIRIS observations.



\subsection{Q1623-BX454}
\label{bx454.sec}

Q1623-BX454 is a narrow-line AGN with both \osix\ $\lambda$1034 and \nfive\ $\lambda$1240 emission in its rest-UV spectrum
that appears to be part of a morphological close pair system.  While Figure \ref{hst_all.fig} indicates Q1623-BX454 itself is compact and featureless in the rest-optical continuum
(GALFIT modeling indicates that it is consistent with an unresolved point source) 
there is also a secondary component located 1.7'' in projection to the southeast of the main body.
Although Q1623-BX454 was not detected in our OSIRIS observations, both components fell within the MOSFIRE slit and were weakly detected in \othree\ emission
with fluxes (assuming a factor of two slit loss correction) of $3 \pm 1 \times 10^{-17}$ erg s$^{-1}$ cm$^{-2}$ and $4 \pm 1 \times 10^{-17}$ erg s$^{-1}$ cm$^{-2}$ for the main (AGN) and
secondary components respectively.

\othree\ emission from the main component is spatially compact (i.e., the observed diameter of 0.9'' is consistent with the observational PSF) with a velocity dispersion of 
 $\sigma = 225 \pm 18$ \kms\ comparable to the other AGN in our sample.  While the secondary component is also spatially unresolved, it 
 has a much lower intrinsic velocity dispersion of $\sigma = 64 \pm 3$ \kms\ that is similar 
 to values observed for typical $z \sim 2$ star forming galaxies.
Since this secondary component is offset by just 167 \kms\ relative to the main body ($z_{\othree} = 2.4181$ vs 2.4200 for the main and secondary components respectively), we conclude that
the AGN is likely interacting with the star-forming companion in a major merger.

Curiously, the relative strength of \othree\ to \Hb\ for the main component is more akin to that observed in galaxies for which star formation
is the primary source of ionizing photons (log($\othree/Hb$) = 0.38).  Coupled with a spectral energy distribution that shows 
effectively zero contribution
from an obscured AGN component (even in the Spitzer $8\micron$ channel, see Figure \ref{sed.fig}), it is unclear to what extent the central
AGN has much effect on the galaxy.  Were it not for the rest-UV emission features such as \osix, \nfive, and \cfour\ that point to excitation by a
hard ionizing spectrum the identification of Q1623-BX454 as an AGN at all would be in doubt.


\subsection{Q1700-MD157}

Q1700-MD157 is a member of the $z = 2.30$ protocluster in the HS 1700+643 field \citep{steidel05} whose \cfour\ width of 1900 \kms\ places it near
our (somewhat arbitrary) classification boundary between narrow-line AGN and broad-line QSO.
In addition to the particularly strong \hetwo\ $\lambda$1640 emission in its rest-UV spectrum, Q1700-MD157 is a
known X-ray source with luminosity $L_{2-10 {\rm keV}} = 4.03^{+0.67}_{-0.58} \times 10^{44}$ erg s$^{-1}$ \citep{dn10}.
Based on the rest-UV morphology (Figure \ref{hst_all.fig}) this source appears to be relatively compact, although it may have a faint companion
that is nearby in projection ($\sim 0.5$'') but of uncertain physical association.
Unfortunately, Q1700-MD157 was not detected in our OSIRIS observations, and has not yet been observed with MOSFIRE.


\subsection{SSA22a-D13}

SSA22a-D13 is a moderately bright (${\cal R} \sim 21$) broad line QSO with strong \lya\ emission at a higher redshift ($z=3.35$)
than the rest of our AGN sample.  Similarly to Q1700-MD157 it is a known X-ray source with 
$L_{2-8 {\rm keV}} = 4.6 \pm 0.7 \times 10^{44}$ erg s$^{-1}$ \citep{lehmer09}.
As illustrated by Figure \ref{d13.fig}
the rest-optical continuum morphology is dominated by the unresolved central QSO and is bright enough to exhibit
a central peak with surrounding Airy ring indistinguishable from stars in the {\it HST} field of view.
The OSIRIS \othree\ image is only marginally
extended with a characteristic
circularized half-light radius $r_{\rm c} = 0.8$ kpc and an integrated velocity dispersion of $192 \pm 13$ \kms.

In this case our OSIRIS obervations successfully detected \Hb\ emission as well, which we find is
comparably bright to  \othree\  ($F = 3.3 \times 10^{-16}$ vs 
$3.7 \times 10^{-16}$ erg s$^{-1}$ cm$^{-2}$ respectively), and traces a spatially
unresolved region with velocity dispersion $\sigma = 300 \pm 50$ \kms.  Although the observed \Hb\ emission is therefore a little broader than the \othree\ component, it is nonetheless similar enough
that it likely traces gas at a similar distance from the central QSO, and any component from the classical broad-line region may be too faint to detect with OSIRIS.





\subsection{Q2343-BX333}

Similarly to Q1623-BX454, Q2343-BX333 is another narrow-line AGN that is not detected in our OSIRIS \Ha\ observations
and weakly detected in \othree\ emission with MOSFIRE (Figure \ref{mosfire1.fig}).
While the HST/F160W continuum image is mildly extended ($r_{\rm c,HST} = 0.9$ kpc; Figure \ref{hst_all.fig})
the MOSFIRE \othree\ image is spatially unresolved (FWHM of 0.56'',  comparable to the 0.57'' observational PSF)
with a velocity dispersion $\sigma = 227 \pm 13$ \kms. 

\subsection{Q2343-BX415}
\label{bx415.sec}

Q2343-BX415 is a $z \sim$ 2.57 QSO that is moderately bright in the optical 
(${\cal R} \sim 20$) but is more luminous for its redshift at 
24 $\mu$m ($f_{\nu, 24\micron} = 196.1 \pm 7.8$ $\mu$Jy) than 
most other AGN or optically selected galaxies
and is comparable in brightness to submillimeter
galaxies (SMGs) at similar redshifts \citep{reddy06,reddy06b}.

As illustrated in Figure \ref{bx415.fig} the morphology of Q2343-BX415 is dominated by the central AGN in our rest-optical imaging, which shows the characteristic structure
of the HST PSF.  In contrast, while the \othree\ emission has a strong central component there is also appreciable flux in a diffuse component that extends a few tenths of an arcsecond
to the northwest and southeast of the core.
This extended emission gives rise to a total source-integrated spectral profile (Figure \ref{bx415.fig}, lower panel) somewhat different for our OSIRIS observations than in previous NIRSPEC
long-slit spectra; while the NIRSPEC profile is relatively symmetric with $\sigma_v = 240 \pm 6$ km s$^{-1}$
the OSIRIS spectrum has an asymmetric blue wing and excess narrow flux near the peak velocity (corresponding to the additional contribution from the extended features missed
by the NIRSPEC observations due to the orientation of the slit)
that produces a source-integrated velocity width $\sigma_v = 220 \pm 2$ km s$^{-1}$.  Clearly, the precise spectral profile depends substantially and systematically on the spatial
regions included in the extraction aperture in a manner belying the formal uncertainties $\sigma_v$; both estimates are somewhat lower than the
central \hetwo\ velocity dispersion $\sigma = 294$ \kms\ measured from the high-resolution ESI spectrum.

Fortunately, we can characterize the properties of the extended emission more completely by 
performing a wavelength-dependent subtraction of the central point source.
In this case, the continuum emission from Q2343-BX415 (which is expected to be dominated 
by unresolved emission from the inner regions of the central AGN)
is sufficiently bright that it provides a real-time cospatial means of assessing the actual LGSAO PSF rather 
than extrapolating from on-axis observations of the tip-tilt reference star.
We therefore constructed a model of the PSF by stacking the continuum wavelengths in our data cube, 
scaled the model by a multiplicative factor at each wavelength,
and subtracted the resulting 3d model of the central point source source from our observations.

As illustrated by Figure \ref{bx415.fig} (middle left panels), we detect substructure in the extended \othree\ emission corresponding
to peaks located about 0.15''
(1.2 kpc) in projection to the northwest (NW) 
and 0.3'' (2.5 kpc) to the southeast (SE) of the central QSO.  
Gaussian fitting to the collapsed spectra of each region indicates that both are nearly at the systemic redshift;
the narrow ($\sigma_{\rm v} = 129 \pm 4$ \kms)
NW feature is redshifted by $23 \pm 4$ \kms while the 
broader ($\sigma_{\rm v} = 195 \pm 8$ \kms)
SW feature is blueshifted by $57 \pm 7$ \kms.
These extended structures have a total diameter of
about 0.8'', or 6.5 kpc.
We conclude that the \othree $\, \lambda5007$ flux of the NW and SE features is $f_{\othree} = 1.8 \pm 0.4 \times 10^{-16}$ \cgs\ and 
2.4 $\pm$ 0.5 $\times 10^{-16}$ \cgs, corresponding to luminosities
$L_{\othree} = 3 \times 10^{43}$ erg s$^{-1}$ and  $4 \times 10^{43}$ erg s$^{-1}$ respectively.\footnote{We assume an 
$E(B-V)=0.26$ based on 
photometric estimates of the UV spectral  slope ($\beta = -1.65$).}


It is unclear whether these features represent gas tracing the gravitational potential of the host galaxy or outflowing
material driven by a central wind, and whether it is ionized by star formation in the host or by the flood of photons
from the central QSO.
Careful subtraction of the central point source in the HST/WFC3 broadband image (J. Anderson; private communication)
suggests the presence of some residual features that represent stellar continuum emission (Figure \ref{bx415.fig}, middle-left
panel); however, these features are inconclusive and
are not cospatial with the extended \othree\ emission.
If the features were due to star formation, we might expect to see a narrow component to the \Ha\ emission in this region
\citep[e.g.,][]{canodiaz12,cresci15,carniani16}.
However, our OSIRIS \Ha\ observations (Figure \ref{bx415.fig}, central panel) showed only an extremely broad component with $\sigma = 1300 \pm 200$ \kms (i.e., tracing gas in
the central unobscured region of the accretion disk), and contained no evidence of kinematically narrow substructures after subtraction of the central point source.
 A tentative line diagnostic assessment is permitted by an upper limit on narrow \Hb\ emission (such as might be expected to arise from star formation in a host galactic disk) in the OSIRIS spectrum, which suggests that
log$(\othree/\Hb) \gtrsim1.0$, consistent with excitation by the hard ionizing spectrum of the central AGN.
Considering the limits on the shape of the ionizing spectrum provided by the strong emission line ratios, the lack of a clear continuum counterpart, and the velocity dispersions
in the residual features (which are roughly twice what is normally observed for star-forming galaxies), 
we favor the interpretation that the NW and SE features represent spatial substructure (and potentially a 
biconical gaseous outflow) within the extended narrow-line region that is
ionized by the strong radiation field of the central AGN.

As discussed by \citet{rix07}, Q2343-BX415 is also
notable for the presence in its spectrum of a metal-poor ($Z \sim 0.2 \, Z_{\odot}$) proximate
damped Ly$\alpha$ absorption system (PDLA) coincident in redshift with the quasar.
The kinematics of this PDLA, as traced by rich rest-UV spectral features, show a complex combination of both
blueshifted low-ionization gas ($\sim 0$ to -350 \kms)
and a redshifted highly ionized gas component ($\sim +100$ to +600 \kms relative to the QSO systemic redshift) which only 
partially covers the continuum source.  As discussed by these authors, 
the presence of excited fine structure absorption lines such as {\textrm{C\,{\sc ii}}}* in the PDLA spectrum implies 
ionization of the PDLA by an intense radiation field longward
of the Lyman limit (suggesting close $\sim 8-37$ kpc physical proximity to the QSO) and
the blueshifted gas (the bulk of which lies at -130 \kms) may therefore represent the wind-driven
outflowing ISM of an as-yet undetected QSO host galaxy.



\begin{figure*}[p!]
\plotone{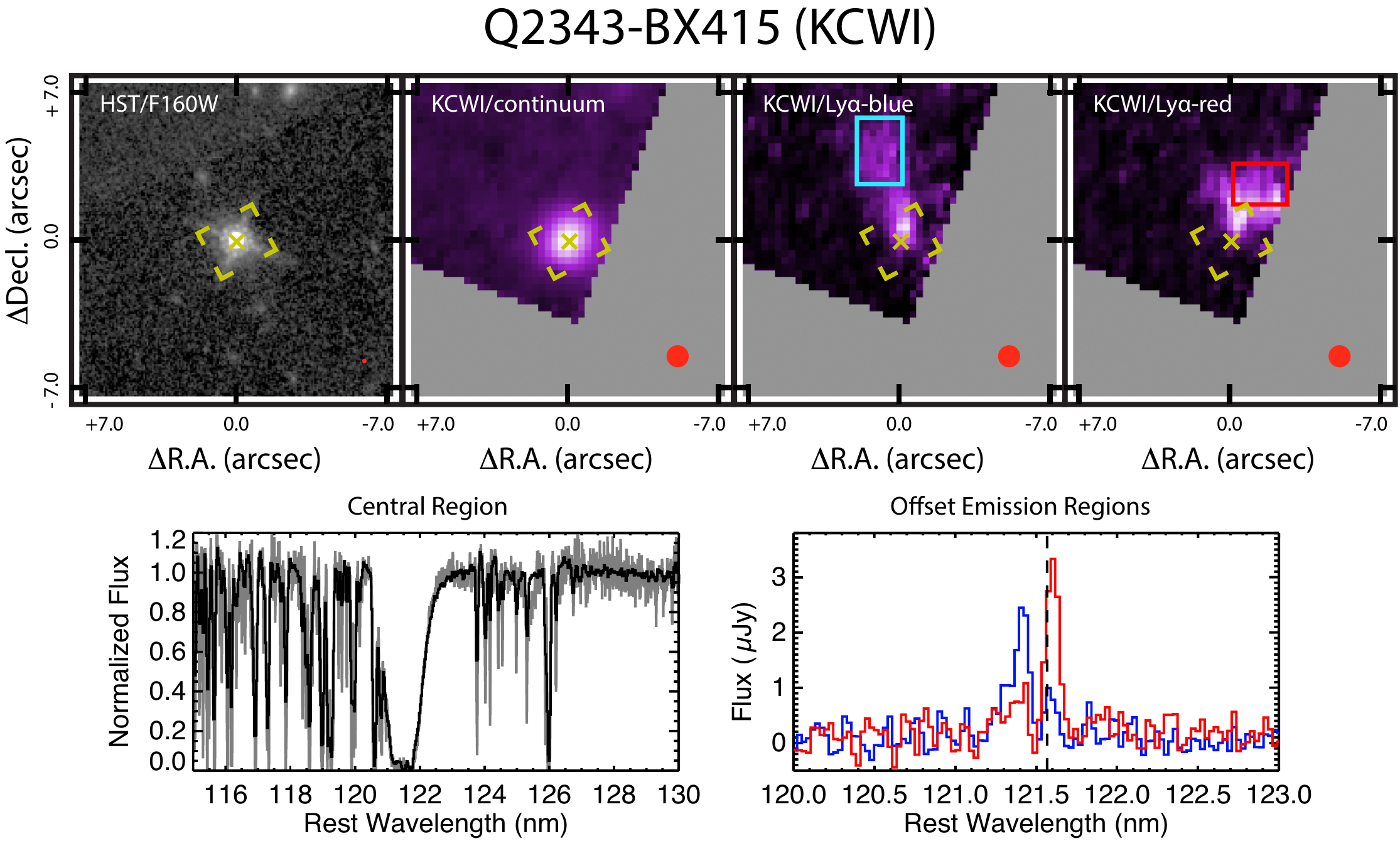}
\caption{
Top row: HST/WFC3 F160W imaging and Keck/KCWI integral-field spectroscopy of Q2343-BX415
illustrating the rest-optical and rest-UV continuum (both shown with logarithmic stretch) and \lya\ emission 
morphology (filled red circles in the lower-right corner of each panel indicate the FWHM of the observational
PSF).
The \lya\ emission is double-peaked with spatially distinct emission regions for the blue (velocity -373 \kms) and red (velocity +84 \kms) components.
Note that the field of view shown in this figure is larger than in previous figures; the dashed yellow box
indicated the field of view of the Keck/OSIRIS observations, while the blue and red boxes indicate the extraction apertures for the blueshifted and redshifted components of the \lya\
emission respectively.  The grey shaded region represents areas outside of the KCWI field of view.
Bottom row, left panel: KCWI continuum-normalized rest-frame spectrum of the central QSO (black line)
along with the higher-resolution Keck/ESI spectrum (grey line) for comparison.  Bottom row, right panel: KCWI
spectra of the offset \lya\ emission regions.  Blue and red lines correspond to the blue and red boxes in the upper panels respectively; the black dashed line indicates the QSO systemic
rest frame of the \lya\ 1215.67 \AA\ transition.
}
\label{bx415lya.fig}
\end{figure*}

With the recent addition of Keck/KCWI integral field spectroscopy we can build upon this picture further.
In Figure \ref{bx415lya.fig} we demonstrate that despite the near-complete resonant absorption of \lya\ photons
by the PDLA there are extended regions of \lya\ beyond the boundaries of the OSIRIS field of view.
This \lya\ emission is double-peaked with components
at 4339.53 \AA\ and 4346.14 \AA\ (corresponding to -373 \kms\ and +84 \kms\ respectively at the systemic redshift of the QSO $z = 2.5741$)
and total flux $F_{\lya} = 7 \times 10^{-17}$ erg s$^{-1}$ cm$^{-2}$ corresponding to a rest-frame
luminosity of $L_{\lya} = 4 \times 10^{42}$ erg s$^{-1}$.
These spectral components are
kinematically narrow ($\sigma \approx 50-100$ \kms\ after accounting for the instrumental line spread function) and spatially distinct;
the blueshifted region has a long filamentary structure reaching up to at least 50 kpc in projection to the north of the QSO, while the redshifted component is more compact and concentrated around
8-30 kpc in projection to the northwest of the QSO (and possibly further since Q2343-BX415
is located at the edge of our KCWI field).
Plausibly, given the similar orientation between the redshifted \othree\ and redshifted \lya\ emission these features may trace different phases within an extended outflowing wind or jet
photoionized by the central QSO.  Likewise, the elongated filamentary morphology of the blueshifted \lya\ feature is consistent with expectations for gas {\it accreting} onto the QSO
through a dynamically cold flow.  However, given the complex resonant nature of the \lya\ emission and its susceptibility to the detailed morphology, kinematics, and covering fraction of the
circumgalactic medium surrounding the QSO both of these structures are difficult to interpret conclusively.



\section{Discussion}
\label{discussion.sec}

 \subsection{Morphology}
 \label{morphology.sec}
 
The role of mergers in producing AGN-mode feedback such as that observed in our sample
is unclear; although such feedback may be triggered by the large quantities of gas funneled onto 
the central AGN during a  major merger, the time scale of the AGN
 mode and the merger-like morphology \citep[e.g.,][]{lotz10} are not necessarily synchronized.
 In our sample of twelve optically faint AGN we find a mixture of 
rest-optical continuum and \othree\ emission-line morphologies; 
one (Q1623-BX454) is a spectroscopically confirmed close pair that may likely merge 
within the next Gyr, two (Q0100-BX164, Q1700-MD157) are morphological pairs whose physical association has not been spectroscopically confirmed, five 
show some evidence for extended continuum and/or \othree\ morphologies, and three 
are isolated systems with no significant visible structure beyond the central point source.  Only in one case (Q0142-BX195) is the morphology clearly `merger-like' with close double
nuclei and extensive tidal tails.

The pair fraction in our sample is therefore in the range of $16 \% - 33 \%$, depending on whether the two apparent morphological pairs are 
genuine physical pairs or not.  This pair fraction is consistent with the pair fraction of the parent star forming galaxy sample
from which our AGN sample was drawn.  In \citet{law12a} we found that the apparent pair fraction (i.e., the number of systems whose rest-optical morphologies show a companion 
within 2'' of uncertain physical association) for $z \sim 2$ star forming galaxies was $23^{+7}_{-6}$ \%; consistent with the 33 \% observed here.  Similarly, in \citet{law15} we demonstrated
using deep spectroscopic observations that $\sim 50$\% of such apparent pairs in the galaxy sample had spectroscopic redshifts consistent with physical association.
We would therefore expect 1.5 genuine spectroscopic pairs in our sample of 12 AGN, consistent with the two such systems observed.
To within the accuracy permitted by low number statistics,  we therefore conclude that 
there is no compelling evidence that the optically faint AGN population is necessarily and uniquely triggered by mergers, similar to previous studies of the optically luminous QSO
population by \citet{mechtley16} and \citet{grogin05} \citep[see also][]{rosario15}.

The most distinguishing morphological feature of our AGN sample is instead that they tend to be
significantly smaller than the star forming sample at comparable stellar mass in both their rest-optical continuum 
\citep[e.g.,][]{nagy11,fs11,law12a} and ionized gas structures \citep[e.g.,][]{law09,fs09,fs18}.  In \citet{law12a}, we found that star forming galaxies with stellar mass
$M_{\ast} = 10^{10} - 10^{11} M_{\odot}$ at $z = 2.0 -2.5$ had mean circularized effective optical continuum radii $\langle r_{c, {\rm HST}} \rangle = 1.84 \pm 0.13$ kpc; in contrast, the sample presented here
has $\langle r_{c, {\rm HST}} \rangle = 1.0$ kpc.  Further, this average is distorted by the lone outlier with an extremely extended stellar envelope (Q0100-BX164); the median AGN in our sample
is unresolved in the optical continuum with $r_{c, {\rm HST}} < 0.5$ kpc.  Such unresolved systems are rare ($< 10$\%) for mass-matched star forming galaxies, but constitute 70\% of the AGN sample.

These smaller sizes are perhaps unsurprising for the broad-line systems in which emission from the central QSO is expected to contribute a significant fraction
of the observed continuum luminosity (although one such QSO actually has the {\it largest} $r_{c, {\rm HST}}$ in our sample).  The narrow-line AGN are less trivially interpreted though since the obscured
central point source should contribute only minimally to the optical continuum light.
Plausibly, the strong correlation between AGN activity and compact size
may support formation mechanisms where the AGN are triggered following dissipative contraction and a compact nuclear starburst \citep[e.g.,][]{kocevski17}.



\subsection{Star Formation within the AGN Host Galaxies}
\label{ratios.sec}

While AGN feedback is generally regarded as a plausible means of truncating the high star formation rates observed in $z \sim 2-3$ galaxies the detailed mechanism and timing of such effects
is uncertain.  Multiple studies suggest, for instance, that increased AGN activity correlates with higher star formation rates as both are fueled by a larger cold gas supply \citep[e.g.,][]{chen13,hickox14}.  AGN-driven feedback however may be substantially delayed with respect to peak star-formation episodes \citep[e.g.,][]{hopkins12} or even exhibit a periodic `flickering' effect \citep[][]{schawin15}.
Likewise, AGN-mode feedback may be both positive and negative, with outflows both suppressing and triggering star formation in different locations within a galaxy \citep[see recent summary by][and references therein]{cresci18}. 

 \cite{azadi17} for instance recently used Keck/MOSFIRE long-slit spectroscopy to study 55 Type II
AGN at $1.4 < z < 3.8$ selected using a combination of X-ray, IR, and optical selection techniques and found no statistically significant
correlation between the presence or absence of an AGN and the observed SFR of the host galaxy compared to a mass-matched
parent sample.  Indeed,
 \citet[][]{az16} found evidence for substantial \Ha-derived star formation rates ($\sim 320 \, M_{\odot}$ yr$^{-1}$) around 16 of 28
hyperluminous Type 1 QSOs using VLT/SINFONI IFU spectroscopy while
 \citet[][]{canodiaz12}, \citet[][]{carniani16}, and \citet[][]{cresci15} described SINFONI observations
of QSOs which appear to show both quenching of star formation in outflow-dominated regions of the galaxy {\it and}
potentially triggered star formation occuring along the edges of the AGN-driven outflow.
Likewise, in observations of five optically faint AGN at a median redshift $z \sim 1.5$
\citet{kakkad16} noted that they could not conclusively determine whether the \othree-derived mass outflow rates
$1- 10 M_{\odot}$ yr$^{-1}$ were powered by the central AGN or by the high SFR ($\sim 100-400 \, M_{\odot}$ yr$^{-1}$)
of the host galaxies.

\begin{figure*}[p!]
\plotone{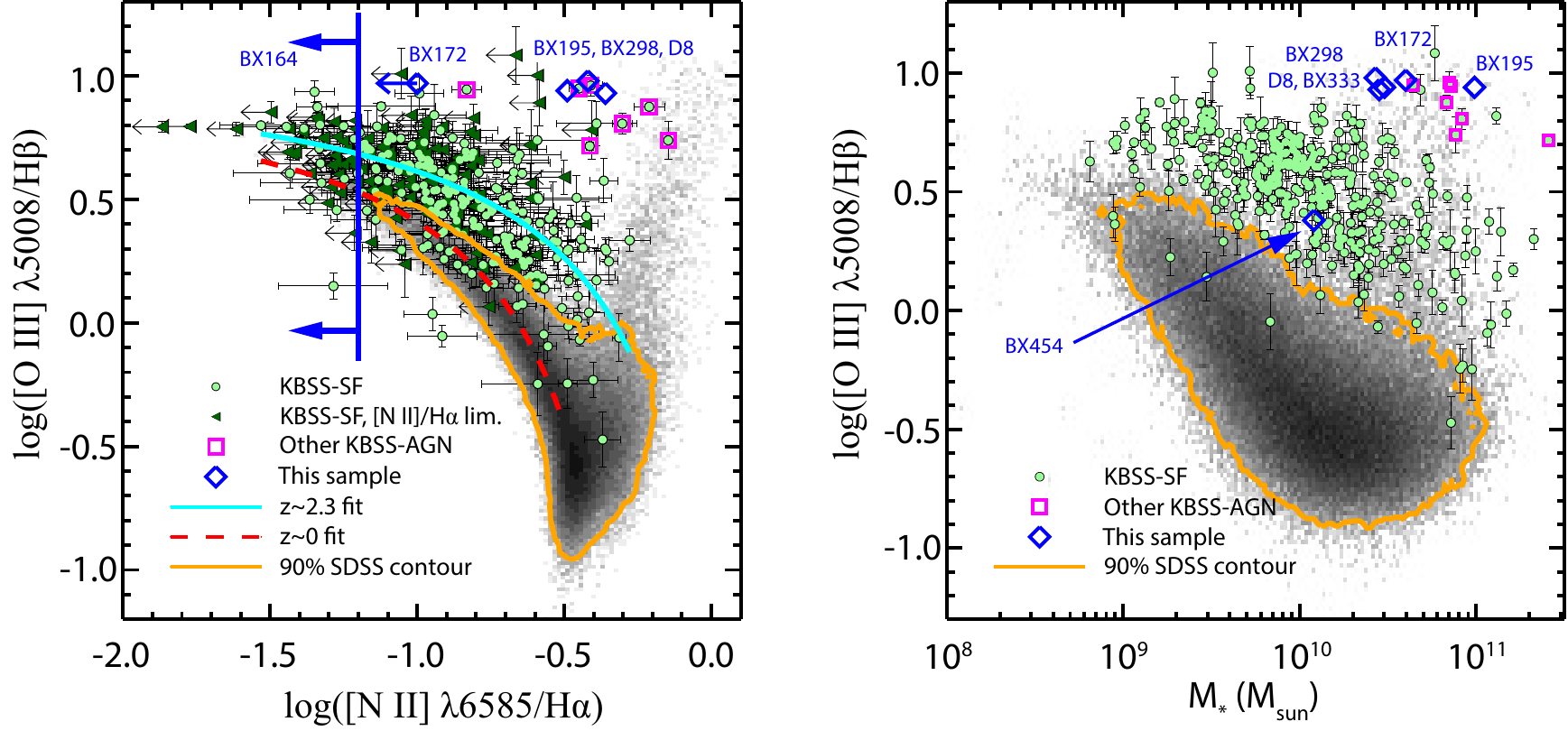}
\caption{N2-BPT diagnostic diagram (left-hand panel) and mass-excitation relation (MEx, right-hand panel) for AGN in the present sample (blue diamonds) compared to the KBSS-MOSFIRE sample of
$z \sim 2$ star forming galaxies and optically faint AGN presented by \citet[][green and magenta symbols respectively]{strom17}.  Each of the blue diamonds is labelled with an abbreviated name.
Shaded greyscale points represent the local relations observed for galaxies in the SDSS.
Except for Q1623-BX454, all of the AGN for which we have been able to measure line ratios lie well above the star forming sequence
at comparable stellar mass.
}
\label{bpt.fig}
\end{figure*}

One major challenge in such efforts to determine the level of star formation present in AGN host galaxies is to discriminate whether the warm gas component is ionized
by photons injected by a central AGN (in which case the size and total luminosity
of the ionized region encodes information about the accretion efficiency of the AGN; see \S \ref{accretion.sec}), 
by photons produced in star forming regions distributed throughout the galaxy, or (likely) through a combination of the two.
A traditional means of determining the {\it dominant} ionization mechanism is provided by the classical strong-line emission diagnostics
\citep[e.g.,][]{bpt81} that can provide an indication of the spectral hardness of the incident radiation field.
As discussed by \citet{steidel14} and \citet{strom17} (and references therein), the entire sequence of $z \sim 2$ star forming galaxies is shifted
in \othree/\Hb\ relative to local galaxies, driven largely by an increase in the hardness of the ionizing
radiation from star forming regions at fixed N/O and O/H ratios (see Figure \ref{bpt.fig}).  
Similar to such previous studies, we find that most of the Type II AGN for which 
multi-band MOSFIRE spectra are available lie significantly off this shifted relation (blue points in Figure \ref{bpt.fig}), indicative of an even-harder ionizing spectrum attributable to an active nucleus.

Following \citet{azadi17} we estimate the possible contribution of star formation to the observed \othree\ flux for our narrow-line AGN for which multi-band spectroscopy is available
by computing the offset distance of the \othree/\Hb\ and \ntwo/\Ha\ ratios
from the modified \citet{kh09} star-formation sequence.  We find that the nebular line
ratios for Q0100-BX172, Q0142-BX195, Q0207-BX298, and Q0821-D8 are offset from this relation by 0.15, 0.39, 0.49, and 0.52 dex, 
suggesting that star formation contributes less than 33\%, 17\%, 12\%, and 11\% respectively to the total observed \othree\ flux.
These fractions are in line with expectations based on our stellar population modeling; as indicated by Table \ref{sed.table}, the star formation rates of the first two
(Q0100-BX172, Q0142-BX195) are $\sim 30 M_{\odot}$ yr$^{-1}$, consistent with the median of a mass-matched KBSS comparison sample of 
$\sim 35 M_{\odot}$ yr$^{-1}$ \citep{hainline12}.  With the aid of our IFU spectroscopy, we can further identify that the star formation in Q0100-BX172 may be occurring in \Ha-bright clumps at
large radius from the center of the galaxy that we detect in our OSIRIS data (\S \ref{bx172.sec}), while the star formation in Q0142-BX195 may be located preferentially in the
extensive gas-rich tidal tails.  Likewise, Q0207-BX298 and Q0821-D8 showed little evidence for star formation in their broadband photometric colors, and have emission line ratios
consistent with this picture.
Even without a measurement of \ntwo/\Ha\
the observed value of {\rm log}\othree/\Hb\ $=0.94$ for Q2343-BX333 is sufficiently 
high for its stellar mass that it is consistent with other KBSS AGN in the mass-excitation relation (MEx; see Figure \ref{bpt.fig}, right-hand panel), suggesting
that it is likewise dominated by the hard ionizing spectrum of the AGN.
In contrast, the low value of {\rm log}\othree/\Hb\ $=0.38$ for Q1623-BX454 places it squarely in the midst of the $z \sim 2$ star forming galaxy sequence (see discussion in \S \ref{bx454.sec}),
suggesting that for this galaxy the observed \othree\ emission may be due to ongoing star formation.


Such line-ratio estimates are more challenging to interpret for the broad-line objects in our sample as it is generally not possible to disentangle the broad
and narrow components of the \Ha\ and \Hb\ emission lines.
However, the complete absence of \othree\ emission in our MOSFIRE spectra for Q0100-BX160 and Q0100-BX164 suggests little to no ongoing star formation\footnote{The nebular metallicity may instead be abnormally high (resulting in low \othree\ excitation), but this possibility seems less likely than the simple absence of significant star formation.};
for reasonable assumptions about the $L_{\othree}$ to $L_{\Ha}$ conversion factor
for star-forming galaxies (see discussion in \S \ref{bx195.sec}) the limit of $L_{\othree} < 2 \times 10^{41}$ erg s$^{-1}$
corresponds to a limit of $< 1 M_{\odot}$ yr$^{-1}$ on the associated star formation rate.

We therefore conclude that the observed nebular emission from our objects is {\it in general} produced primarily by gas heated by
the hard ionizing spectrum of the central AGN, although some individual objects (Q0100-BX172, Q0142-BX195, Q1623-BX454) show potential evidence for contributions from star forming regions that may be far from the galaxy centers.  It is difficult to speculate about the role that the AGN may have played in
actively suppressing star formation activity in our sample however, given the different timescales relevant for the two.

 \subsection{Gas-phase kinematics}
\label{kinematics.sec}
 
 As discussed in \S \ref{results.sec} for individual objects, the \othree\ velocity dispersions of the AGN sample ($\langle \sigma \rangle \sim 230$ \kms) are significantly larger than
those  observed in  typical  $z \sim 2$ star forming galaxies of comparable mass
\citep[$\sigma_{\rm gas} = 60-80$ \kms, see, e.g.,][]{fs09,law09,simons17}.
While there can be a large difference between the {\it localized} gas velocity dispersion $\sigma_{\rm gas}$ and the {\it source-integrated} velocity dispersion $\sigma_{\rm tot}$
for star forming galaxies with substantial velocity gradients, this distinction appears to be minimal for the AGN sample presented here which appear to be compact and show little evidence
for spatially resolved velocity shear.  Indeed, in the instance where a rotational
component is most well-justified (Q0207-BX298) the central velocity dispersion is sufficiently large that the rotational signature broadens it only minimally.

In part, the higher velocity dispersions are likely tied to the compact nature of the sample, especially given that \othree\ emission in the AGN sample traces the distribution of NLR clouds 
rather than star-forming H II regions.
Following simple virial arguments, at fixed stellar mass compact galaxies such as these that fall below
the mass-radius relation would be expected to have higher effective velocity dispersions, and similar trends have been observed for other 
compact galaxy samples at similar redshifts \citep[e.g.,][]{barro14, vd15}.

At the same time, the ionized gas kinematics may also be intrinsically tied to a greater prevalence of outflowing winds driven partly by the central AGN.
\citet{harrison16} for instance used the KMOS IFU to observe a sample of X-ray selected type II AGN with 
$L_{\othree} \sim 10^{41} - 10^{43}$ erg s$^{-1}$ at
redshifts $z = 0.6 - 1.7$, and found that large ionized gas velocities 
(defined as $W_{80} > 600$ \kms, or $\sigma \gtrsim 234$ \kms\ for a single gaussian component)
indicative of outflows are ten times more prevalent in AGN host galaxies compared to a matched sample
of star forming galaxies.
Similarly, \citet{talia17} observed that their sample of 79 X-ray selected AGN at $z = 1.7 - 4.6$ 
($L_{\rm X} \sim 10^{42} - 10^{45}$ erg s$^{-1}$) had interstellar absorption lines blueshifted from the systemic redshift (as defined by stellar photospheric features)
by 950 \kms\ on average compared to about 150 \kms\ for their mass-matched star forming galaxy comparison sample \citep[see also][]{hainline11}.

Although our KBSS sample of optically faint AGN tends to be less X-ray luminous than these two samples
\citep[see, e.g.,][]{reddy06} we nonetheless find similar evidence for enhanced outflow velocities in the ionized gas.
 Five of our AGN (Q0142-BX195A, GOODSN-BMZ1384, Q1623-BX454, Q2343-BX333, Q2343-BX415) fulfill the  $W_{80} > 600$ \kms\
 criterion, while three (Q0100-BX172, Q0207-BX298, and GOODSN-BMZ1384\footnote{Using the MOSFIRE spectrum
 presented by \cite{leung17}}) exhibit clear blue/red wings in their \othree\ profiles
 that reach to $\pm 1000$ \kms.
 Such features are often observed in local Seyfert galaxies \citep[e.g.,][]{heckman81,greene05,rice06} and interpreted as evidence of
outflowing gas, and similar blue wings
have been observed previously in the \othree\ profiles of  bright
AGN at $z \sim 2$ \citep[e.g.,][]{cresci15,kakkad16}.
 The incidence of such broad wings in our sample is 25\% (i.e., 3/12), which is similar to the value
 of 19\%  recently obtained by \citet{leung17}
 from the MOSDEF survey and roughly 10 times more common than in the star-forming galaxy population.

\subsection{Size of the ENLR}
\label{enlrsize.sec}


Given that the majority of the observed \othree\ emission appears to be excited by ionizing radiation from the central AGN (\S \ref{ratios.sec}),
we assess whether the size of the extended emission line region (ENLR) is consistent with expectations from active galaxies in the nearby
universe.
Historically, the size of the ENLR has been observed to increase with the amount of ionizing radiation roughly
proportional to  the square root of the
total \othree\ luminosity \citep[e.g.,][]{bennert02,schmitt03}.  As noted by \citet{greene11} and \citet{hainline14} however, the radii measured depend strongly
on the effective depth of a given observation, and efforts to compare different studies must therefore account for both 
the limiting surface brightness of the observations
and cosmological surface brightness dimming.  \citet{liu13} and \citet{hainline14} therefore measure $r_{\rm iso}$, the isophotal radius at which 
the \othree\ surface brightness decreases to
\begin{equation}
\Sigma_{\rm iso} = \Sigma_0 \frac{1}{(1+z)^4}
\end{equation}
where $\Sigma_{0} = 10^{-15}$ \cgs\ arcsec$^{-2}$.   \citet{liu13} find that
observations of low-redshift radio-quiet quasars and Seyfert II galaxies can be well described by the relation

\begin{equation}
\textrm{log} \, r_{\rm iso} = (0.250 \pm 0.018) \, \textrm{log} \, L_{\othree} + (3.746 \pm 0.028)
\end{equation}
where $r_{\rm iso}$ and $L_{\othree}$ are given in units of pc and $10^{42}$ erg s$^{-1}$ respectively.

At the median redshift of our sample ($z \sim 2.4$), cosmological dimming reduces $\Sigma_{\rm iso}$
to $7 \times 10^{-18}$ \cgs arcsec$^{-2}$; far fainter than the typical limiting surface brightness of our OSIRIS observations ($3 \times 10^{-16}$ \cgs\ arcsec$^{-2}$;
see \S \ref{osiris.dq.sec}).  It is therefore not possible to compare the sizes that we measure with OSIRIS for our high-redshift AGN to these low-redshift
relations directly given the factor $\sim 40$ difference in effective depth.

Instead, we estimate $r_{\rm iso}$ by modeling the observed \othree\ radial profile of our galaxies and extrapolating the profile
to the necessary surface brightness.  For each AGN in our OSIRIS sample (and for both AGN in the pair for Q0142-BX195; note that each fitted component represents
only the cores and not the obvious tidal features) we use 
GALFIT 3.0 \citep{peng02,peng10} to model the \othree\ surface brightness maps (Figs. \ref{bx172.fig} - \ref{bx415.fig})
with a two-dimensional Sersic profile convolved with the reference PSF\footnote{Derived from the continuum emission for Q2343-BX415, and the
on-axis tip-tilt star observations for all other galaxies.}  for each galaxy \citep[see, e.g.,][for further details]{law12a}.  The resulting Sersic profile
is given by:

\begin{equation}
\Sigma(r) = \Sigma_e {\rm exp} \left[ - \kappa \left( \left( \frac{r}{r_e}\right) ^{1/n} -1 \right) \right]
\label{sersic.eqn}
\end{equation}

where $r_e$ is the effective radius, $n$ the concentration parameter, and $\kappa$ is related to $n$ such that half of the total flux in contained within $r_e$.
We adopt the approximation \citep{ps97} that 

\begin{equation}
\kappa = 2n - \frac{1}{3} + 0.009876/n
\label{kappa.eqn}
\end{equation}

which is valid within the range $n = 1-4$ that we are considering.  The normalization factor $\Sigma_e$ can be computed \citep[][see their Eqn 7]{peng02} as:

\begin{equation}
\Sigma_e = F_{\othree} \left( 2 \pi r_e^2 e^{\kappa} n \kappa^{-2n} \Gamma(2n) q \right)^{-1}
\label{sigmae.eqn}
\end{equation}
where $F_{\othree}$ is the total observed source flux in units of erg s$^{-1}$ cm$^{-2}$, $\Gamma$ is the complete gamma function,
and $q$ is the axis ratio.

It is then possible to rearrange Eqn. \ref{sersic.eqn} to give an exact expression for the isophotal radius $r_{\rm iso}$ as:
\begin{equation}
r_{\rm iso} = r_e \left[ -\left(\frac{1}{\kappa}\right) \, {\rm ln}\left( \frac{\Sigma_{0}}{\Sigma_{\rm e} (1+z)^4} \right) + 1 \right]^n
\label{finalr.eqn}
\end{equation}
where $\Sigma_{0} = 10^{-15}$ erg s$^{-1}$ cm$^{-2}$ and
 $\Sigma_e$ and $\kappa$ are given by Eqns. \ref{sigmae.eqn} and \ref{kappa.eqn} respectively.

As indicated by Table \ref{results.table}, the PSF-deconvolved effective radii range from 0.54 arcsec down to $\leq 0.06$ arcsec (we adopt a
threshhold of 0.4 times the FWHM of the observational PSF as an upper limit on the size of unresolved sources).  After extrapolation to
a cosmology-corrected surface brightness of $10^{-15}$ erg s$^{-1}$ cm$^{-2}$ using Eqn. \ref{finalr.eqn}, our estimated sizes
$r_{\rm iso}$ range from $<4$ - 17 kpc (0.5'' - 2.1''; Table \ref{results.table}).
The uncertainties inherent in this approach are difficult to quantify as they rely on an extrpolation of the observed
surface brightness profile to levels substantially below the observational limit of the data.  Similarly, they
rely critically on proper deconvolution of the observational PSF \citep[see, e.g.,][]{husemann16} and neglect the known filamentary structure
of ENLR observed in nearby systems \citep[e.g.,][]{bennert02} and in our own sources.

Nonetheless, we find that our estimated sizes based on extrapolation of the OSIRIS IFU spectroscopy agree well with size estimates
based on our MOSFIRE slit spectroscopy.   As discussed in \S \ref{osiris.dq.sec}, the $3\sigma$ limiting surface brightness of the MOSFIRE spectra 
is about $5 \times 10^{-18}$ erg s$^{-1}$ cm$^{-2}$ arcsec$^{-2}$, which is well matched to the surface brightness at which we need to make our measurements.
We therefore estimate $r_{\rm iso}$ from the MOSFIRE spectroscopy by simply measuring the radius along the slit at which the observed \othree\ profile
is equal to the appropriate value at the redshift of each galaxy and subtracting off the known observational seeing in quadrature.\footnote{For the Q2343-BX415
NIRSPEC observation we estimate the observational seeing as the FWHM of the continuum emission from the QSO.}
We tabulate these sizes in Table \ref{mosfire.table}, providing upper limits in the two cases for which the observed \othree\ FWHM was consistent with the
observational PSF, and noting that some other values may be underestimated in cases (e.g., Q0821-D8, Q2343-BX415) for which the spectrograph slit
was misaligned with the axis of greatest \othree\ extent.

\begin{figure}
\epsscale{1.2}
\plotone{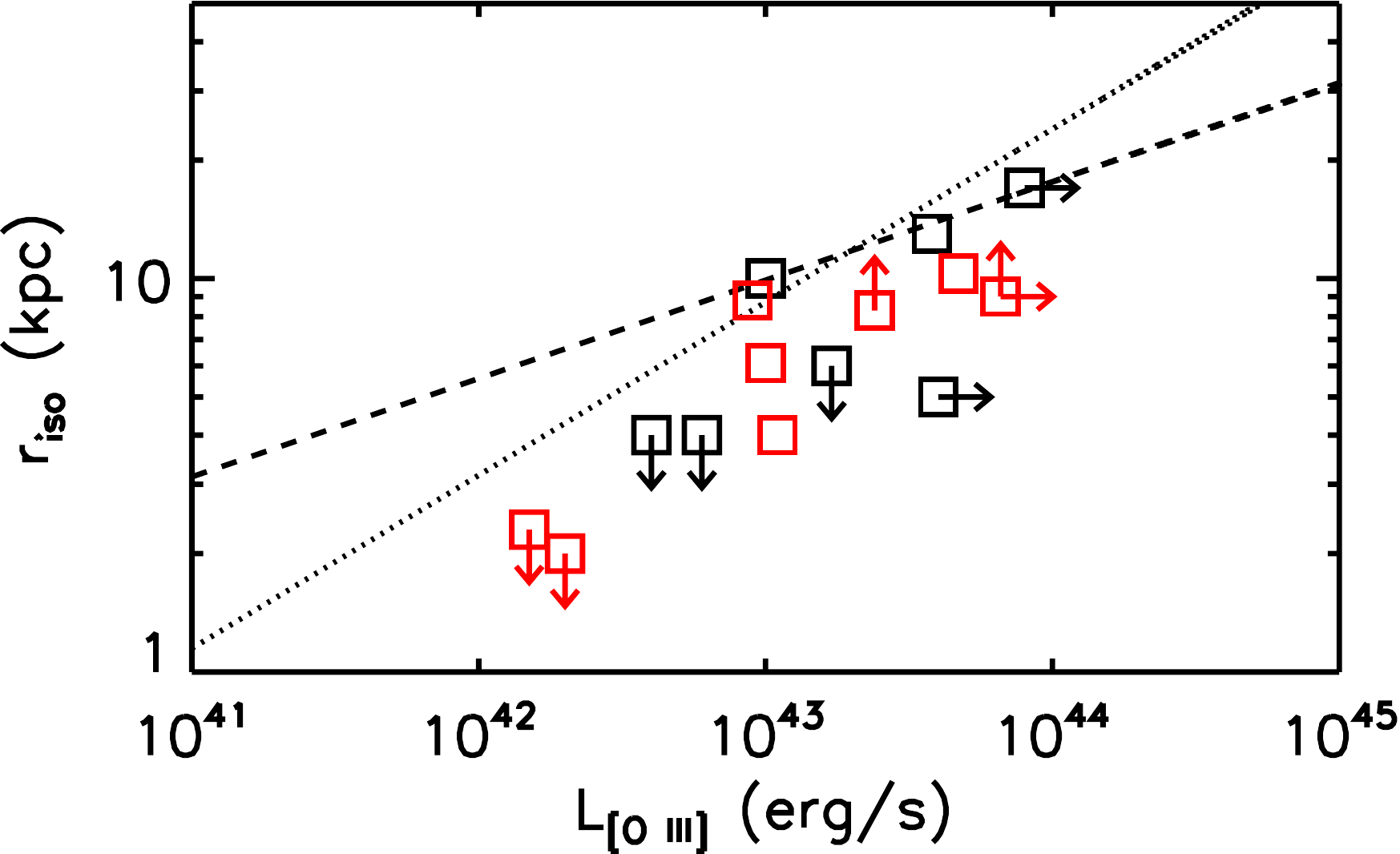}
\caption{Isophotal radius of the ENLR for galaxies in our sample (open boxes) at a surface brightness of $\Sigma_{\rm iso} = 10^{-15}$ erg s$^{-1}$ cm$^{-2}$ 
(corrected for cosmological dimming) as a function of the total \othree\ luminosity $L_{\othree}$.  Black/red symbols represent Keck/OSIRIS and
Keck/MOSFIRE measurements respectively.  Downward-pointing arrows indicate upper limits on unresolved
sources, rightward-pointing arrows indicate Type I QSOs for which no dust correction has been made in computing $L_{\othree}$.  The dashed line indicates
the local relation defined by \citet{liu13}; for comparison we also show the steeper relation of \citet[][dotted line]{husemann14}.
}
\label{lrplot.fig}
\end{figure}

As indicated by Figure \ref{lrplot.fig} our OSIRIS and MOSFIRE measurements are consistent with each other and clearly demonstrate a radius --- luminosity
relation\footnote{Although it is unsurprising to find a relation between $L_{\othree}$ and $r_{\rm iso}$ given that the latter depends on $f_{\othree}$ through the $\Sigma_e$
normalization at fixed $r_e$, the observed relation cannot be entirely due to this effect for reasonable choices of the Sersic parameter $n$.}
akin to that observed locally for radio-quiet quasars and Seyfert II galaxies, albeit potentially offset to a factor $\sim 2$ smaller radii at a given
luminosity (although such a normalization difference may simply represent a systematic error in our method of determining radii).  The slope of our
relation is also somewhat steeper than that of \citet[][$\alpha=0.25$]{liu13} and more akin to that derived by \citet[][$\alpha = 0.52$]{bennert02} and
 \citet[][$\alpha =0.44$]{husemann14}.
 Under the assumption that the extended NLR is produced in a series of optically thick clouds in pressure equilibrium 
 with ionization parameter $U = \Phi/(4 \pi r^2 n_e c)$
(where $n_e$ is the electron density and the ionizing photon production rate $\Phi$ is proportional to the bolometric luminosity of the AGN $L_{\rm bol}$), we note that our result
 $r \sim L_{\othree}^{0.5}$ is consistent with the simple interpretation of a constant ionization parameter
 within the NLR \citep[see, e.g.,][and references therein]{dz18}.


\subsection{Black Hole Accretion Rates}
\label{accretion.sec}

Finally, we consider whether the AGN in our sample are radiating at close to their
theoretical (Eddington) limit at which radiation pressure balances the gravitational force from the central black hole
for the simplified case of spherical accretion.
We define the Eddington ratio $\lambda$ as
\begin{equation}
\lambda = \frac{L_{\rm bol}}{L_{\rm Edd}}
\end {equation}
where $L_{\rm bol}$ is the bolometric luminosity of the AGN and
\begin{equation}
L_{\rm Edd} = 1.3 \times 10^{38} \, (M_{\rm BH}/M_{\odot}) \, {\rm erg \, s}^{-1}
\end{equation}
is the Eddington luminosity and $M_{\rm BH}$ is the mass of the central black hole in solar units.

\subsubsection{Broad-line QSOs}
\label{edd_broad.sec}

For our type I broad line QSOs the standard AGN unification paradigm suggests that we observe the central accretion disk directly,
and it is therefore possible to estimate the black hole mass (and thus the Eddington accretion limit) using observations of the disk continum and broad emission lines.  
Assuming that the gas in this region is virialized, the velocity width of spectral emission features such as \cfour\ and \Ha\ should trace the gravitational potential of
the disk, and along with measurements of the size of the region can provide estimates of the total central mass.
Such relations have been well calibrated in the local universe for a variety of spectral lines using reverberation mapping techniques; we follow
\citet{mr16} in adopting:

\begin{equation}
M_{\rm BH,\cfour} = 2 \times 10^6 \left(\frac{L_{1450}}{10^{44} \, \textrm{erg/s}}\right)^{0.588} \left(\frac{\textrm{FWHM}(\cfour)}{1000 \, \textrm{km/s}}\right)^2 M_{\odot}
\label{methodc4.eqn}
\end{equation}

and

\begin{equation}
M_{\rm BH,\Ha} = 6.0 \times 10^6 \left(\frac{L_{5000}}{10^{44} \, \textrm{erg/s}}\right)^{0.569} \left(\frac{\textrm{FWHM}(\Ha)}{1000 \, \textrm{km/s}}\right)^2 M_{\odot}
\label{methodha.eqn}
\end{equation}
where FWHM(\cfour) and FWHM(\Ha) are the measured FWHM of the \cfour\ $\lambda 1550$ and \Ha\ lines
in our rest-UV and rest-optical spectroscopy respectively, and $L_{1450}$ and $L_{5000}$ are
the rest-uv and rest-optical continuum luminosities.\footnote{$L_{5000} = \nu L_{\nu} =
4 \pi D_{\rm L}^2 \nu f_{\nu}$ where $D_{\rm L}$ is the luminosity distance, $\nu$ is the observed frequency,
and $f_{\nu}$ is the observed frame flux density.}
As indicated by Table \ref{eddington.table} the black hole masses derived using Eqns. \ref{methodc4.eqn} and \ref{methodha.eqn}
from our rest-UV
and rest-optical measurements agree to within a factor of two and range from $\sim
10^8$ to $\sim 10^9 M_{\odot}$.

Further, we assume that the observed optical continuum of these systems is dominated by the radiation from the central accretion disk, and can therefore be used
to estimate the total bolometric luminosity of the AGN.  
Following \citet{heckman04} and \citet{marconi04} we assume that the total bolometric luminosity
of these broad-line systems is related to the continuum luminosity as

\begin{equation}
L_{\rm bol} = 10.9 \times L_{5000}
\end{equation}

on average for a typical Type I spectral energy distribution.  As indicated by Table \ref{eddington.table}, the value $L_{\rm bol} = 6.8 \times 10^{12} L_{\odot}$ 
derived by this method
for Q2343-BX415 is in reasonable agreement with the total uncorrected UV + IR luminosity ($L_{\rm UV} = 2 \times 10^{12} L_{\odot}$, $L_{\rm IR} = 8 \times 10^{12} L_{\odot}$)
derived from our rest-UV and Spitzer 24$\mu$m observations.
As indicated by Table \ref{eddington.table}, we find that the broad-line objects in our sample
are currently radiating within a factor of a few of their Eddington luminosity,
ranging from $\sim 1/5$ Eddington to slightly super-Eddington.

\begin{deluxetable*}{lcccccccc}
\tablecolumns{9}
\tablewidth{0pc}
\tabletypesize{\scriptsize}
\tablecaption{Black Hole Accretion Properties}
\tablehead{
\colhead{Name} & 
\colhead{log$\left(\frac{M_{\rm BH,st}}{M_{\odot}}\right)$\tablenotemark{a}} & 
\colhead{log$\left(\frac{M_{\rm BH,\cfour}}{M_{\odot}}\right)$\tablenotemark{b}} & 
\colhead{log$\left(\frac{M_{\rm BH,\Ha}}{M_{\odot}}\right)$\tablenotemark{c}} & 
\colhead{log$\left(\frac{L_{\rm Edd}}{L_{\odot}}\right)$} & 
\colhead{log$\left(\frac{L_{5000}}{L_{\odot}}\right)$\tablenotemark{d}} & 
\colhead{$L_{5000}/L_{\othree}$\tablenotemark{e}} & 
\colhead{log$\left(\frac{L_{\rm Bol}}{L_{\odot}}\right)$\tablenotemark{f}} & 
\colhead{$\lambda$\tablenotemark{g}}
}
\startdata
\multicolumn{9}{c}{Broad-line QSOs}\\
\hline
Q0100-BX160 & ... & 7.34 & 7.20 & 11.80 & 10.70 & $>1000$ & 11.74 & 0.9 \\
Q0100-BX164 & ... & 8.26 & 7.91 & 12.62 & 10.95 & $>1700$ & 12.00 & 0.2\\
SSA22a-D13 & ... & 8.32 & ... & 12.85 & 11.93 & 83 & 12.97 &  1.3 \\
Q2343-BX415 & ... & 8.84 & 9.04 & 13.47 & 11.79 & 30 & 12.83 & 0.2 \\
\hline
\multicolumn{9}{c}{Narrow-line AGN}\\
\hline
Q0100-BX172 & 7.90 & ... & ... & 12.43  &  ... & ... & 12.78 & 2.2 \\
Q0142-BX195A & 8.00\tablenotemark{h} & ... & ... & 12.53 &  ... & ... & 12.43 & 0.8 \\
Q0142-BX195B & 8.00\tablenotemark{h} & ... & ... &  12.53 &  ... & ... & 12.20 & 0.5 \\
Q0207-BX298 & 7.71 &  ... & ... & 12.24 &  ... & ... & 12.20 & 0.9 \\
Q0821-D8 & 7.75 &  ... & ... & 12.28 &  ... & ... & 12.94 & 4.6 \\
GOODSN-BMZ1384 & 7.84 & ... & ... & 12.37 &  ... & ... & 11.80 & 0.3 \\
Q1623-BX454 & 7.38 &  ... & ... & 11.91 &  ... & ... & 11.37  & 0.3 \\
Q1700-MD157 & 7.98 &  ... & ... & 12.51 &  ... & ... &  ... & ... \\
Q2343-BX333 & 7.78 & ... & ... & 12.31 &  ... & ... & 11.50 & 0.2 
\enddata
\tablenotetext{a}{Black hole mass for Type II AGN estimated as a fraction of the stellar mass $M_{\rm BH,st} = 0.002 M_{\ast}$.}
\tablenotetext{b}{Black hole mass for Type I AGN estimated from rest-UV continuum luminosity and \cfour\ line width (Eqn. \ref{methodc4.eqn}).}
\tablenotetext{c}{Black hole mass for Type I AGN estimated from 5000 \AA\ continuum luminosity and \Ha\ line width (Eqn. \ref{methodha.eqn}).}
\tablenotetext{d}{$L_{5000} = \nu L_{\nu}$ at $\lambda = 5000$ \AA\ rest frame based on broadband photometry for broad-line QSOs.}
\tablenotetext{e}{Ratio between observed \othree\ luminosity and $5000$ \AA\ luminosity (both uncorrected for dust) from broadband photometry for broad line QSOs.}
\tablenotetext{f}{Total bolometric luminosity estimated from dust-corrected \othree\ luminosity (for type II AGN) or $L_{5000}$ (for type 1 AGN).  Estimate uses OSIRIS-derived $L_{\othree}$ if available, and otherwise uses the MOSFIRE-derived value.}
\tablenotetext{g}{Eddington ratio $\lambda = L_{\rm bol} / L_{\rm Edd}$.}
\tablenotetext{h}{Stellar mass estimate is based on photometry that does not resolve the two components; estimated mass is divided equally based on comparable rest-optical continuum magnitudes.}
\label{eddington.table}
\end{deluxetable*}

We note, however, that we would have found a drastically different result had we instead used $L_{\othree}$ as a proxy for the total AGN luminosity.  Since the majority of the gas in the narrow-line region is photoionized by the central AGN (based on the observed emission line ratios) we could also have chosen to adopt the
\cite{heckman04} relation that $L_{5000}/L_{\othree} = 320$ on average for type I AGN in the Sloan Digital Sky Survey, 
and thereby derive $L_{\rm bol} = 3500 L_{\othree}$.
However, as indicated by Table
\ref{eddington.table}, $L_{5000}/L_{\othree}$ for our four broad-line systems varies from $\sim 30$ to $> 1000$ for Q0100-BX160 and Q0100-BX164
(for which we did not detect any \othree\ emission with MOSFIRE).  This range is significantly
larger than the $1\sigma$ dispersion about the mean of 0.34 dex found at lower redshifts by \citet{heckman04} and \citet{zakamska03}.
Clearly, some physical mechanism is suppressing the \othree\ flux in many of our faint QSOs \citep[and in many hyperluminous QSOs as well; see, e.g.,][]{ts12}.\footnote{Note that this effect should
not affect our conclusions regarding the typical size of the ENLR, since the relevant QSOs had no \othree\ detection and hence no \othree\ size measurement.}

Comparing to the \citet{bg92} library of optical spectra from the Palomar Bright Quasar survey, we note that our weak or absent \othree\ emission is similar to QSOs with large
negative values along their eigenvector 1, which is primarily dominated by the anticorrelation between \othree\ and \fetwo\ strength.
As discussed by \citet{bm92} and \citet{turnshek97}, such \othree -weak systems often exhibit broad absorption lines in their rest-UV spectra and have stronger \nfive\
emission.  Indeed, while our rest-UV spectra of Q0100-BX160 and Q0100-BX164 do not show broad absorption features Q0100-BX160 has particularly strong \nfive\ $\lambda 1240$, 
and our rest-optical spectra are similar to those of 
BAL QSOs PG 0043+039 and PG 1700+518 \citep[see also][]{pettini85}.
These similarities suggest that a possible explanation for the absence of \othree\ in Q0100-BX160 and Q0100-BX164 may be that an increase in the broad-line cloud covering
fraction in these systems shields ionizing radiation from escaping to the narrow-line region.  While we might expect the Eddington ratios to be correspondingly higher for such objects (corresponding to high density gas approaching the Eddington accretion limit), we see no evidence
that the Eddington ratios for these objects is any higher that in the other broad-line systems with more `normal' \othree\ luminosities.

%
%

\subsubsection{Narrow-line AGN}
\label{edd_narrow.sec}

In Type II systems such as our narrow-line AGN sample, the obscuration of the central accretion disk means that we must estimate the properties of the central black hole
using more indirect means.
Following  \citet{aird12} and \citet{azadi17} we assume 
that the black hole mass is proportional to the total galactic stellar mass derived from our stellar population modeling via the relation
\begin{equation}
M_{\rm BH,st} = 0.002 \,  M_{\ast}
\label{methodst.eqn}
\end{equation}
This approach yields black hole masses for our sample ranging from $\sim 10^7$ to $\sim 10^8 M_{\odot}$ (see Table \ref{eddington.table}).

Since the central accretion disk is obscured at rest-optical and NIR wavelengths in these systems, the observed continuum luminosity traces the galactic stellar population
and is therefore not a good proxy for the bolometric luminosity of the accretion disk.  However, the observed \othree\ emission is still primarily ionized by the AGN
and arises in extended kiloparsec-scale regions, and therefore might still be a proxy for the total AGN luminosity (although c.f. discussion in \S \ref{edd_broad.sec}).
As discussed in \S \ref{sed.sec}, we estimate the dust-corrected \othree\ luminosity $L_{\othree}$ using the reddening values derived from SED fitting,
and further follow \citet{kh09} and \citet{azadi17} \citep[see also][]{lamastra09} in assuming that the bolometric luminosity is related to the 
dust-corrected \othree\ luminosity as $L_{\rm bol} = 600 \, L_{\othree}$ based on local calibrators.


As indicated by Table \ref{eddington.table}, we find that
the majority of our narrow-line systems are also radiating within a factor of a few of their Eddington luminosity,
consistent with prior estimates based on the rest-UV luminosities of the KBSS optically-faint AGN sample \citep{steidel02}.
As for the broad-line systems, however, our 
estimates of the Eddington ratio depend on the validity of a variety of assumptions.  

First, our estimates of the Eddington luminosity depend upon the total masses derived from stellar population modeling and our
assumption that the $z=0$ black hole --- bulge mass relation holds at $z \sim 2$.
Both observational evidence \citep[e.g.,][]{greene10,ts12} and hydrodynamical simulations \citep[e.g.,][]{barber16} suggest however that black holes in the young universe 
may be overmassive relative to their host galaxies compared to the $z = 0$ relation, possibly by as much as a factor of ten.  Such an effect may lead us to substantially underestimate
the maximal Eddington luminosity of our AGN, and hence overestimate their Eddington ratios.  At the same time, while the compact sizes of our galaxies in the rest-optical continuum
suggests that the total stellar mass may be a reasonable proxy for the bulge mass we might nonetheless expect that this difference results in a systematic {\it underestimate} of the Eddington ratio.

Second, our calculation of the bolometric luminosity from the dust-corrected \othree\ luminosity may also be biased.
Notably, the highest Eddington ratio object in our sample (Q0821-D8; $\lambda = 4.6$) drops to $\lambda = 2.3$
if we use the observed Balmer decrement ($F_{\Ha}/F_{\Hb} = 2.9$) instead of the SED-derived extinction to dust-correct the \othree\ luminosity.  Likewise, 
our assumption that the bolometric luminosity is related to the \othree\ luminosity by a simple multiplicative factor with reasonably low dispersion
is known to be incorrect for the broad-line objects in our sample and can 
produce dramatically different results from other methods of estimating the bolometric luminosity. 
\citet{hainline12} for instance estimate the AGN luminosity by integrating the AGN component of their fit to the observed SED, and for a sample of $z \sim 2$ AGN similar to our own
find estimates of $L_{\rm bol}$ that are systematically smaller by an average of 1.5 dex.  Applying this technique to our sample, we likewise find estimates of the AGN bolometric luminosity
that are factors of ten or more lower than obtained using the $L_{\othree}$ method, although individual values are highly uncertain given the extrapolation of the AGN SED far beyond their small
contributions to the observed photometric data.
Unsurprisingly, this difference in methods drives the disagreement between the nearly Eddington-limited accretion rates that we derive using the $L_{\othree}$ method
($\langle{\rm log} \, \lambda \rangle = -0.2$) and the substantially sub-Eddington rates discussed by
\citet[][$\langle{\rm log} \, \lambda \rangle = -1.5$]{hainline12}.
Our results are instead closer to those of \citet{azadi17} who use similar methods to ours in calculating $\lambda$.  In particular,
their IR-selected AGN sample (which has a stellar mass distribution most closely matched to our UV-selected sample) has median log $\lambda = -0.4$ similar to our own.
Although \citet{azadi17} find that their
X-ray and optical line-ratio selected samples have lower Eddington ratios (median $\lambda = -0.92$ and -1.25 respectively), it is likely that these samples represent quite different AGN
populations as their stellar masses are about 0.5 dex more massive on average than in our own UV-selected sample.

\section{Conclusions}
\label{conclusions.sec}

We have presented Keck/OSIRIS and Keck/MOSFIRE spatially resolved spectroscopy of \othree\ and \Ha\ emission from a sample of optically faint AGN
drawn from the KBSS  at a median redshift of $z \sim 2.4$.
By combining our observations with a wide range of ancillary photometric and spectroscopic data (ranging from the optical to mid-infrared, and including
rest-UV integral field spectroscopy) we
have analyzed the morphology, outflow kinematics, and ionization mechanisms of twelve optically faint AGN
(four dominated by broad permitted line emission, and eight with only narrow-line emission) 
whose rest-UV colors and magnitudes are generally representative of the parent star forming galaxy population.
Our major conclusions can be summarized as follows:
\begin{itemize}

\item The rest-optical continuum morphologies of the AGN show no particular hallmarks of merging activity 
compared to the parent star forming sample from which they were drawn
(see \S \ref{morphology.sec}).  While there are some clear major-mergers in our sample (e.g., Q0142-BX195, which
exhibits a spectacular double active nucleus with extended tidal features) the overall fraction of such systems is consistent with the fraction observed in a mass-matched
sample of ordinary star forming galaxies at comparable epochs.  The AGN are, however, substantially more compact, consistent with 
formation mechanisms in which they are triggered following dissipative contraction and a compact nuclear starburst.

\item Unlike recent results in the literature for brighter AGN, we do not detect widespread evidence for residual or triggered star formation
in the AGN host galaxies or their outflowing winds (\S \ref{ratios.sec}).  Where measurable, our diagnostic line ratios are 
consistent with photoionization by the hard spectrum of the central AGN, and our measured velocity dispersions are substantially higher
than observed in typical star forming galaxies at similar epochs and inferred stellar masses.  The exceptions to this general result are 
Q0100-BX172 (which shows some
evidence for widely distributed knots of compact \Ha\ emission) and Q0142-BX195 (for which one of the two tidal tails
has  substantial \othree\ emission that may be due to photons produced in a local burst of star formation within the tail).

\item Consistent with multiple recent studies and observations in the nearby universe, we find that our AGN are accreting at
nearly Eddington-limited rates (subject to numerous systematic uncertainties; see \S \ref{accretion.sec})
and preferentially drive strong outflows into the surrounding intergalatic medium.  These outflows exhibit
both blueshifted and redshifted wings on the \othree\ emission line profiles that reach up to 1000 \kms\ relative to the systemic redshift (\S \ref{kinematics.sec}).

\item Local-universe scaling relations between \othree\ and bolometric or 5000-\AA\ luminosity for optically faint broad-line AGN appear to break down
for the redshift $z \sim 2$ sample, with a substantial fraction of such systems undetected in \othree\ in our deep MOSFIRE observations, implying
$L_{5000}/L_{\othree} > 1700$ (\S \ref{edd_broad.sec}).

\item Based on the observed \othree\ surface brightness profile, the size of the extended narrow-line region for optically faint $z \sim 2$ AGN appears
consistent with local scaling relations for active nuclei where $r \sim L^{0.5}$ (see \S \ref{enlrsize.sec}).
Our sample of AGN fall a factor of two
below the local relation on average however, reflecting either an evolution in the typical ionization profile of the narrow-line clouds or simply the 
difficulty in extending the local isophotal surface brightness threshhold to high redshifts
given the strong $(1+z)^4$ cosmological surface brightness dimming.

\end{itemize}

\acknowledgements
The authors thank Alice Shapley, Dawn Erb, Naveen Reddy, and Max Pettini for their participation in early KBSS observations, particularly the
near-IR and {\it Spitzer} mid-IR photometry and the Keck/LRIS spectroscopic campaign.
DRL also thanks Alice Shapley for assistance with early OSIRIS observations and 
Jay Anderson for assistance with PSF subtraction of the Q2343-BX415 HST imaging, and appreciates
productive conversations with Tim Heckman, Nadia Zakamska, and the STScI AGN Reading Group.
This work includes observations taken by the CANDELS Multi-Cycle Treasury Program 
and the 3D-HST Treasury Program (GO 12177 and 12328)
with the NASA/ESA HST, which is operated by the Association of Universities for Research in Astronomy, Inc., under NASA contract NAS5-26555.
Finally, we wish to extend thanks to those of Hawaiian ancestry on whose sacred mountain
we are privileged to be guests.

\end{document}